\documentclass[reqno,11pt]{amsart}
\usepackage{amsmath, amssymb, amsthm,amsfonts} 
\usepackage[english]{babel}
\usepackage{bbm,bm}
\usepackage{graphicx}
\usepackage{url}
\usepackage{epstopdf}
\usepackage[a4paper,bindingoffset=0.2cm,left=1cm,right=1cm,top=2.5cm,bottom=2cm,footskip=.8cm]{geometry}
\usepackage{subcaption}
\usepackage{textcomp}
\usepackage[]{algorithm2e}
\usepackage{multicol}
\usepackage{multirow}
\usepackage{wrapfig}
\usepackage{diagbox}
\usepackage{floatrow}
\usepackage{xfrac}
\newfloatcommand{capbtabbox}{table}[][\FBwidth]

\usepackage{tikz}
\usetikzlibrary{arrows,chains,fit,automata,positioning,calc,shapes,decorations.pathreplacing,decorations.markings,shapes.misc,petri,topaths,backgrounds}
\usepackage{pgfplots}
\pgfplotsset{compat=newest}
\usetikzlibrary{plotmarks}
\usepackage{grffile}
\newlength\figureheight
  \newlength\figurewidth
\pgfplotsset{%
    tick label style={font=\scriptsize},
    label style={font=\footnotesize},
    legend style={font=\footnotesize},
         every axis plot/.append style={very thick}
}
\tikzset{
    ncbar angle/.initial=90,
    ncbar/.style={
        to path=(\tikztostart)
        -- ($(\tikztostart)!#1!\pgfkeysvalueof{/tikz/ncbar angle}:(\tikztotarget)$)
        -- ($(\tikztotarget)!($(\tikztostart)!#1!\pgfkeysvalueof{/tikz/ncbar angle}:(\tikztotarget)$)!\pgfkeysvalueof{/tikz/ncbar angle}:(\tikztostart)$)
        -- (\tikztotarget)
    },
    ncbar/.default=1cm,
}

\usepackage{rotating}
\usepackage{amsbsy,enumerate}
\usepackage{graphicx}
\usepackage{comment}
\usepackage{mathrsfs} 
\usepackage{tikzscale}
\usepackage{xcolor}
\usepackage[comma]{natbib}

\newcommand{\vb}{\vspace{5mm}}

\allowdisplaybreaks

\makeatletter
\newtheorem*{rep@theorem}{\rep@title}
\newcommand{\newreptheorem}[2]{%
\newenvironment{rep#1}[1]{%
 \def\rep@title{#2 \ref{##1}}%
 \begin{rep@theorem}}%
 {\end{rep@theorem}}}
\makeatother

\makeatletter
\renewcommand*\env@matrix[1][\arraystretch]{%
  \edef\arraystretch{#1}%
  \hskip -\arraycolsep
  \let\@ifnextchar\new@ifnextchar
  \array{*\c@MaxMatrixCols c}}
\makeatother

\newtheorem{theorem}{Theorem}
\newreptheorem{theorem}{Theorem}

\newtheorem{remark}{Remark}

\newreptheorem{proposition}{Proposition}

\begin{document}

\title[Estimating Probability Distributions of Travel Times by
Fitting a Markovian Velocity Model]
{Estimating Probability Distributions of Travel Times \\ by
Fitting a Markovian Velocity Model}

\author{Nikki Levering, Marko Boon, and Michel Mandjes}

\maketitle

\begin{abstract}
To improve the routing decisions of individual drivers 
and the management policies designed by traffic operators,
one needs reliable estimates of travel time distributions. 
Since congestion caused by both recurrent patterns (e.g., rush hours) 
and non-recurrent events (e.g., traffic incidents) leads to
potentially substantial delays in highway travel times, 
we focus on a framework capable of incorporating 
both effects.
To this end, we propose to work with the Markovian velocity model,
based on an environmental background process that tracks both random and
(semi-)predictable events affecting the vehicle speeds in a highway network.
We show how to operationalize this flexible data-driven model
in order to obtain the travel time distribution
for a vehicle departing at a known
day and time to traverse a given path.
Specifically, we detail how to structure the 
background process and set the speed levels
corresponding to the different states of this process.
First, for the inclusion of non-recurrent events, we 
study incident data to 
describe the random durations of the
incident and inter-incident times.
Both of them 
depend on the time of day, but we identify
periods in which they can be considered time-independent.
Second, for an estimation of the speed patterns
in both incident
and inter-incident regime, loop detector data
for each of the identified periods is studied.
In numerical examples 
that use road network
detector data of the Dutch highway network, we obtain
the travel time distribution estimates that
arise under different traffic regimes,
and illustrate the advantages compared to
traditional travel-time prediction methods.

\vb


\vb

\noindent
{\sc Affiliations.} 
{NL and MM are with the Korteweg-de Vries Institute for Mathematics, University of Amsterdam, Amsterdam. MB is with the Department of Mathematics and Computer Science, Eindhoven University of Technology, Eindhoven. MB and MM are also with E{\sc urandom}, Eindhoven, and MM with Amsterdam Business School, University of Amsterdam, Amsterdam.}
This research project is partly funded by the NWO Gravitation project N{\sc etworks}, grant number 024.002.003. Date: \today. 

\vb

\noindent
{\sc Copyright notice.}
This work has been submitted to the IEEE for possible publication. Copyright may be transferred without notice, after which this version may no longer be accessible.

\end{abstract}

\newpage

\section{Introduction}

{\sc Motivation and short method description} \\
Accurate and efficient estimation of travel time
distributions is 
needed to help
individual travelers make well-informed routing decisions.
Moreover, traffic operators use information about travel time distributions
for the design of optimal policies for traffic management.
Consequently, a reliable description of these travel times
may lead to reductions of delays, economic costs, 
and $\text{CO}_2$ emissions.
Considering vehicle trip times, one should distinguish 
between so-called {\it recurrent 
congestion} (i.e., near-periodic effects, such as
congestion during peak hour) and
{\it non-recurrent congestion} (which is inherently less predictable, 
e.g.\ covering congestion
due to incidents) potentially contributing to substantial
delays. When aiming at describing travel time distributions, it is therefore essential to include
both effects.
The objective of  this paper is to develop, in the context of a highway network, a framework for capturing
the impact of recurrent and non-recurrent congestion. 

Since the origins of recurrent congestion are of an essentially periodic nature,
it follows a highly predictable pattern. This can be inferred from
velocity data, often available through the loop detectors 
or speed cameras present in traffic networks.
In contrast, incidents are considerably less 
predictable, both in terms of location and severity. 
Our work focuses on developing a description of the
randomness regarding such incidents, and a quantification of
their impact on highway travel times, in a model that 
in addition takes the recurrent, near-periodic effects into account.
In our approach we incorporate events that directly 
impact the velocities at which the vehicles {\it can} drive 
(which we refer to as `driveable speed levels'), 
thus ignoring second-order effects 
that are caused by the driving style that 
individual drivers may have.

We demonstrate our approach by studying traffic 
data from the Dutch highway 
network, which we use to get a handle on the (random) 
incident lengths, inter-incident times and corresponding
driveable speed levels.
The analysis employs loop detector data, in combination with
a database of registered incidents. While 
we use the Dutch data in our `proof of concept', 
our techniques can be applied to any highway network
for which similar data sets are available.
Importantly, with the results of the
analysis, we operationalize 
the {\it Markovian velocity model}~({\sc mvm}), as was introduced
in \cite{Levering2022AConditions}.
This stochastic model uses an environmental
background process to track both recurrent and
non-recurrent events affecting vehicle speeds
in a road network, and outputs,
given its departure day and time,
a description of the travel time of
a vehicle traversing a specific path. 
Notably, 
this description yields an accurate proxy for the 
{\it distribution} of this travel time, rather than 
just a `point estimate', thereby providing
insight into the impact of the random effects discussed above.
With increasing recognition of travel time
reliability as an important performance measure,
such distributions serve as input for 
a line of routing
studies that take the risk-averseness 
of users into account (see e.g.\ \cite{Fosgerau2010TheReliability}).

A main reason for our choice to use the {\sc mvm} to model travel
time distributions lies in the fact that it is `velocity oriented' and data-driven.
In addition, it is remarkably
flexible in terms of its capacity to include the various sources of travel time fluctuations.
Concretely, the proposed framework 
transparently captures the causes of recurrent as well as non-recurrent congestion 
in a single model. Indeed, 
we can incorporate near-deterministic events 
(such as the onset of the rush hour, or the duration of the rush hour), 
as well as events of an intrinsically more random nature (such as incidents).
Moreover, the underlying mechanism is rich enough to allow for
correlation between the speeds on different 
segments in the network, present due to e.g.\ spillback
and rubbernecking after incidents.
Lastly, as is demonstrated in this paper, despite the flexibility the model offers,
it can be made operational with relatively low complexity,
which makes it directly useful for practical purposes.
One such practical application of the {\sc mvm}
is studied in \cite{Levering2022AConditions}, who employ the model
in an optimal routing context, in which an individual vehicle
wishes to minimize its expected travel time between a given origin
and destination.

{\sc Literature review} \\
Since incidents potentially have a dramatic impact on 
highway travel times, their duration has been
studied extensively. Some early works describe the
randomness of incident durations by the lognormal
distribution \citep{Garib1997EstimatingDelays, Giuliano1989IncidentFreeway, Sullivan1997NEWDELAYS}. 
Reviews of more recent
studies \citep{Chung2012AnalyticalAnalysis,Li2018OverviewPrediction}
reveal that, besides the
lognormal distribution, the log-logistic and 
Weibull distribution are frequently found 
to describe the random duration of
incidents well. 
\cite{Li2018OverviewPrediction} distinguish
between the \textit{analysis} and \textit{prediction} of traffic incidents.
On the one hand, analysis studies have the 
objective to determine
which factors have a significant impact on the incident
duration. Types of factors that are found to affect
the duration include environmental conditions,
the characteristics of the incident, and
traffic flow conditions. 
On the other hand, prediction studies have the objective to forecast the duration of a current incident.
Reviewed prediction methods are e.g.\ regression models, artificial neural networks, 
and hazard-based duration models.

An important remark is that the applicability of the
incident distributions reported in 
\cite{Chung2012AnalyticalAnalysis} and \cite{Li2018OverviewPrediction}
is limited for the description of {\it future} incidents.
First, these studies do not consider the incident rate (i.e., the rate at which a new incident occurs),
and are therefore unable to describe the time
until a future incident.
Second, incident durations are often modeled
under various configurations of explanatory
variables, but information regarding the values of 
these factors may only be available if an incident
has actually occurred, or even only if an incident
has elapsed for a certain period of time
(e.g.\ number of involved vehicles, number of closed lanes).
Thus, since these prediction methods are
only useful once this information becomes available,
they are of limited use for predicting the duration of future
incidents or incidents that just occurred. The latter case
is also studied by \cite{Khattak1995ADuration} and \cite{Ghosh2019DynamicSet}, who
account for the chronological availability of information
by presenting a time-sequential
prediction method that updates the prediction when
new information becomes available.
Importantly, for current incidents, the {\sc mvm} model that we advocate in this paper offers the same flexibility,
while additionally being able to describe
the time until \emph{and} the duration 
of future incidents. 

The Markovian velocity model of \cite{Levering2022AConditions}
includes current incidents, 
future incidents, and daily patterns in the travel time distribution, 
by capturing 
the effect of recurrent and non-recurrent events
on highway speeds. 
In contrast to most travel time prediction
methods (see e.g.\ the summaries in \cite{vanHinsbergen2007ShortModels},
\cite{Vlahogianni2014Short-termGoing} and
\cite{Qiu2021MachineAnalyses}), the
model outputs a distribution of the travel time
rather than a point estimate, so as to provide insight
in the impact of traffic incidents.
Moreover, it is one of the few models that directly
models the impact of traffic incidents on travel times.
That is, most studies regarding travel time distributions
focus solely on daily patterns, and describe the travel
time distribution for different periods of day.
Examples of such recent studies include 
\cite{Chalumuri2014ModellingJapan,Guessous2014EstimatingConditions,Chen2019DataPrediction}
and \cite{Chen2020AnalyzingData}.
In the data-driven
models of \cite{Filipovska2021EstimationCorrelations},
both daily patterns and incidents \textit{are} incorporated, but,
as they investigate general path travel time distributions,
there is no focus towards incidents that are in
the network at the moment of the vehicle's departure.

To the best of our knowledge, 
two of the few studies that assess
the impact of current incidents on
the travel time are the regression models presented by
\cite{TavassoliHojati2016ModellingReliability}
and \cite{Javid2018AData}.
However, the impact
of the incident on the travel time is only
indirectly quantified, as the models use a 
travel time reliability measure as response variable.
Similarly, \cite{Xie2020DeepPrediction} do not directly
investigate travel times during incidents, but focus on
the problem of incident-driven speed prediction,
and, to this end, propose the use of a specific
graph convolutional network.
\cite{Miller2012MiningImpact} do consider 
the incident-induced delay directly, but only predict the
incident impact as a class variable.

The Markov model of \cite{Levering2022AConditions} 
is a natural extension of 
the travel time models presented by 
\cite{Kim2005OptimalInformation, Kim2005StateInformation, Yeon2008TravelChains} and
\cite{Guner2012DynamicInformation}.
These studies use a Markovian background process
to model the daily recurrent patterns and
let the state of the process on a link directly impact
the travel time on this link, thereby neglecting spatial correlation.
In contrast, besides daily patterns, 
the background process in \cite{Levering2022AConditions} is used to model more
complex traffic events, such as traffic incidents,
and recognizes the correlation between link travel times.
Moreover, similar
to \cite{Kharoufeh2004DerivingProcesses},
the travel time adheres the FIFO-property,
as the state of the continuous background process
impacts the \textit{driveable vehicle speed} instead of
the travel time.

{\sc Main contributions} \\
The contributions of this paper are twofold. In the first place,
we demonstrate how traffic data can be used
to obtain an accurate description of the randomness
of incidents. This description involves the incidents' frequency, 
duration and impact on traffic velocities.
These turn out to typically depend on the time of day and day of week, but one 
can deal with these fluctuations by working with
periods over which these effects are essentially constant.
As mentioned above, the data study concerns the Dutch highway network, 
but our techniques extend to all
highway networks for which incident
and speed data is available.

In the second place, we show how to include both the randomness
of incidents and the recurrent traffic patterns, so as to obtain
the travel time distribution of a vehicle traversing a path through the network.
Concretely, we use the results from the data study to
operationalize the {\sc mvm}, thus tracking  both 
(near-deterministic) time-dependent and intrinsically random events.
We explicitly show how incidents and daily patterns can be incorporated
into the background process, and how their effect
on vehicle speeds must be chosen to accurately reflect
their impact on the travel time of a vehicle.
By doing so, we provide traffic management centers and individual drivers
with a transparent modelling framework, which they can easily
calibrate to their needs, so as to obtain travel time distribution estimates.

\mbox{{\sc Paper organization}}
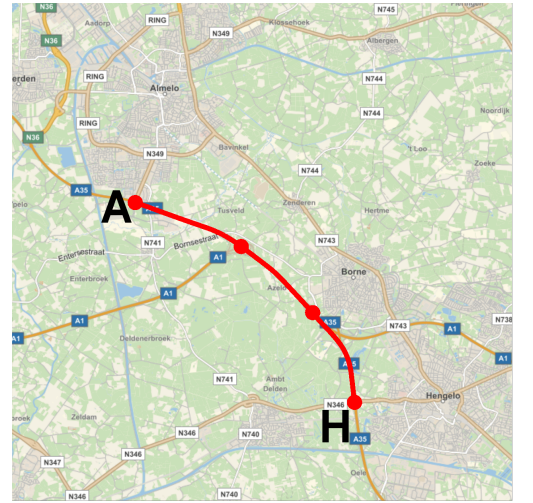
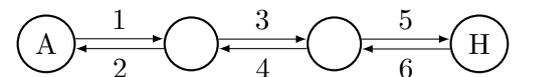
\begin{wrapfigure}{r}{.35\textwidth}
    \begin{minipage}{\linewidth}
    \centering\captionsetup[subfigure]{justification=centering}
    \includegraphics[width=\linewidth]{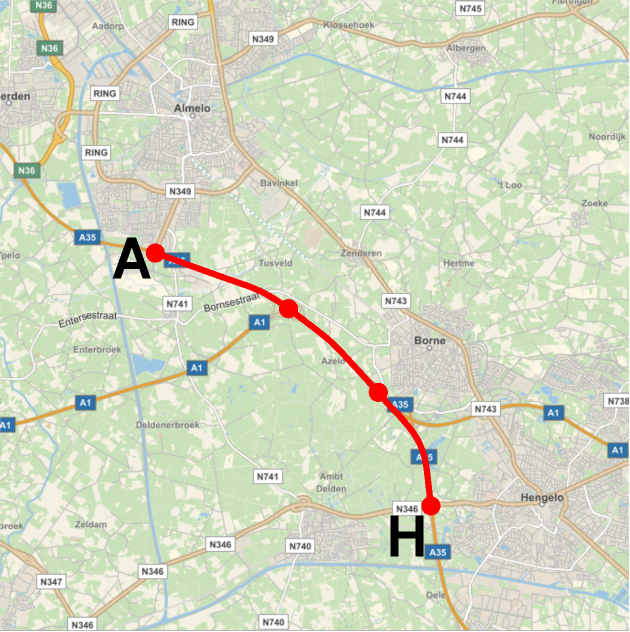}
    \subcaption{A35 highway path between Almelo and Hengelo.}
    \label{fig:mapAH}\par\vfill
    \begin{tikzpicture}[visnode/.style={circle, draw=black, thick, minimum size=7mm}]
\node[visnode] (A) {A};
\node[visnode,right=3em of A] (B)  {};
\node[visnode,right=3em of B] (C)  {};
\node[visnode,right=3em of C] (D)  {H};
\draw[-latex] (A.10)-- node [above] {1} (A.10-|B.west);
\draw[-latex] (B.-170)-- node [below] {2} (B.-170-|A.east);
\draw[-latex] (B.10)-- node [above] {3} (B.10-|C.west);
\draw[-latex] (C.-170)-- node [below] {4} (C.-170-|B.east);
\draw[-latex] (C.10)-- node [above] {5} (C.10-|D.west);
\draw[-latex] (D.-170)-- node [below] {6} (D.-170-|C.east);
\end{tikzpicture}
    \subcaption{Graphical scheme. Nodes represent
ramps, the indexed directed links the highway parts between
the ramps.}
    \label{fig:graphicalAH}
\end{minipage}
\caption{Path Almelo-Hengelo}\label{fig:AH}
\end{wrapfigure}
The {\sc mvm} 
is compactly described in Section~\ref{sec:MVM}. Section~\ref{sec:dataanalysis} starts
with a description of the considered network and corresponding
traffic data, to then continue with an analysis 
of this data for arcs in both the incident and non-incident setting.
Section~4 details how the observations from the
data analysis can be used to operationalize the
{\sc mvm}. Numerical examples of the resulting travel time distributions
are given in Section~5. 
Section~6 contains concluding remarks.

\section{Markovian velocity model} \label{sec:MVM}
The main objective of this section is to briefly describe the Markovian velocity model ({\sc mvm}),
as developed in \cite{Levering2022AConditions},
which we propose to use for the prediction of travel time
distributions. 
As touched upon in the introduction, our choice for the {\sc mvm} stems from the
transparency and extreme flexibility it offers for the modeling of
both recurrent and non-recurrent traffic events, 
in terms of their frequency, duration,
and their impact
on network velocities.
Indeed, from a data study of 
the Dutch highway network (Section~\ref{sec:dataanalysis}),
it will become
apparent that the {\sc mvm} framework
is well capable of describing
the vehicle speeds in this network.
It is important to note that we focus on the (recurrent and non-recurrent) events that directly affect
the speed levels the vehicles {\it can} drive at. 
This concretely means that our approach does not 
incorporate velocity-impacting effects due to e.g.\
the individual drivers' heterogeneity in driving style.

The {\sc mvm} uses a
{\it background process}, or {\it environment process}, to track the near-deterministic
and random events affecting the vehicle speeds 
in the road network.
Section~\ref{subsec:modelingexample}
provides an illustrative example
for the structure of this background process,
when considering a vehicle traversing a path in 
a small network 
with typical traffic events. 
A more detailed description
of the mathematical framework of the {\sc mvm}
is presented in Section~\ref{subsec:mathmodel}.
In Section \ref{subsec:fittingevents} we discuss general principles underlying statistically fitting traffic events. 

\subsection{Modeling example} \label{subsec:modelingexample}~\\
Consider a vehicle that intends to 
traverse the A35 highway in the Netherlands
between Almelo and Hengelo (Figure~\ref{fig:mapAH}), 
and that enters this highway in Almelo at
a moment at which there is no reported incident.
Figure~\ref{fig:graphicalAH} shows the
graph that corresponds to the path, 
with each node representing
a ramp on the highway,
and each link representing
the highway part between two of these ramps.
As in the rest of this paper, we are interested
in the travel time of the vehicle planning
to traverse the path, given the vehicle enters 
this path at a specific day and time.

Observe that the travel time a vehicle experiences
can be inferred from the speeds
the vehicle is able to drive.
On the considered part of the
A35 highway, the maximum speed
as set by the Dutch government equals 100~km/h.
However, the attained speed on the arcs is not necessarily
this maximum.
In reality, events such as rush hour and
traffic incidents lead to fluctuations in vehicle speeds.
To model these effects,
the {\sc mvm} introduces a
background process that tracks the events 
affecting the speeds.

A prominent source of speed variability is
formed by randomly occurring traffic incidents.
To model these incidents,
we let $\{X_i(t),t\!\geq\!0\}$ be a 
continuous-time Markov
process that records whether there is an 
incident on link $i$ at time $t$.
Specifically, we choose
\begin{align*}
    X_{i}(t) &= 
    \begin{cases}
    1 &\quad \quad \text{if there is an incident on arc $i$ at time $t$,} \\
    2 &\quad \quad \text{otherwise},
    \end{cases}
    \quad \text{ and } \quad 
    Q_i = \begin{bmatrix}
    -\alpha_i & \alpha_i \\ \beta_i & -\beta_i
    \end{bmatrix},
\end{align*}
with $Q_i$ the transition rate matrix
of $X_i(t)$ with transition rates $\alpha_i, \beta_i > 0$.
Thus, in this example, we observe that $X_i(t)$ is a process
that cyclically switches between an exponentially distributed time (with mean
$1/\alpha_i$) during which there is an incident on arc $i$, and an exponentially
distributed incident-free time (with mean $1/\beta_i$).
We let, for different links $i$ and $j$, the processes $X_i(t)$ and $X_j(t)$ evolve
independently.
Then, the {\it superimposed} process $B(t) := (X_1(t),\dots,X_6(t))$ is a Markovian background process
recording the incidents in the network of Figure~\ref{fig:graphicalAH},
having a state space of dimension $2^6$.
Setting $t\!=\!0$ as the time the vehicle
enters the A35 highway at Almelo,
we know $B(0)\!=\!(2,2,\dots,2)$, as
there were no reported incidents at that time.

\begin{figure}[ht]
\centering
\begin{tikzpicture}[visnode/.style={circle, draw=black, thick, minimum size=7mm}]
\node[visnode] (A) {\tiny A};
\node[visnode,right=3em of A] (B)  {};
\node[visnode,right=3em of B] (C)  {};
\node[visnode,right=3em of C] (D)  {\tiny H};
\node[visnode,right=3em of D] (A1) {\tiny A};
\node[visnode,right=3em of A1] (B1)  {};
\node[visnode,right=3em of B1] (C1)  {};
\node[visnode,right=3em of C1] (D1)  {\tiny H};
\node[draw=none,left=0.5em of A] (Z1) {\small (a)};
\node[draw=none,left=0.5em of A1] (Z2) {\small (b)};
\draw[-latex] (A.10)-- node [above] {100} (A.10-|B.west);
\draw[-latex] (B.-170)-- node [below] {100} (B.-170-|A.east);
\draw[-latex] (B.10)-- node [above] {100} (B.10-|C.west);
\draw[-latex] (C.-170)-- node [below] {100} (C.-170-|B.east);
\draw[-latex] (C.10)-- node [above] {100} (C.10-|D.west);
\draw[-latex] (D.-170)-- node [below] {100} (D.-170-|C.east);
\draw[-latex] (A1.10)-- node [above] {60} (A1.10-|B1.west);
\draw[-latex] (B1.-170)-- node [below] {100} (B1.-170-|A1.east);
\draw[-latex] (B1.10)-- node [above] {30} (B1.10-|C1.west);
\draw[-latex] (C1.-170)-- node [below] {80} (C1.-170-|B1.east);
\draw[-latex] (C1.10)-- node [above] {100} (C1.10-|D1.west);
\draw[-latex] (D1.-170)-- node [below] {100} (D1.-170-|C1.east);
\end{tikzpicture}
\caption{Example of speed levels (in km/h) on the A35 highway
between Almelo ({\sc A}) and Hengelo ({\sc H}), in case of (a) no incidents,
(b) an incident on link 3. The latter shows the effect of spillback
and rubbernecking.
}
\label{fig:graphicalAH1}
\end{figure}
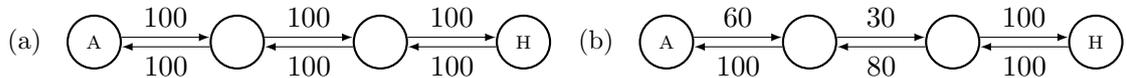

Now, if, during the traversal of
the path between Almelo and Hengelo, 
an incident would
occur at one of the links, naturally, 
the speed level on this link is affected
by the incident.
An important observation
is that the impact on 
vehicle speeds may not be limited
to the incident link itself.
That is, potential spillback
and rubbernecking effects
may lead to speed reductions
on upstream links
or the link on the other side
of the barrier as well. 
To reflect this
dependence, 
we let
the velocity on an arc be
determined by the {\it complete}
state of the background process~$B(t)$.
Concretely, if $B(t)$ is in state 
$s\!\in\!\{1,2\}^6$, the vehicle
speed at arc~$i$ equals $v_i(s)$.
Indeed, modeling the speeds in this fashion,
the speed on link~$i$ does not solely
depend on $X_i(t)$, but is allowed to
depend on all $X_j(t),~j\!=\!1,\dots,6$.
Thus, if we would want to capture the
typical traffic behavior during an
incident at link~3, with estimated
speed levels as displayed 
in Figure~\ref{fig:graphicalAH1},
we could simply set the speeds
in state $s = (2,2,1,2,2,2)$
equal to 
$(v_1(s),v_2(s),v_3(s),v_4(s),v_5(s),v_6(s))\!=\!(60,100,30,80,100,100)$.

\begin{figure}[ht]
    \centering
\begin{tikzpicture}[shorten >=1pt,node distance=2cm,on grid,auto]
    \node[state,inner sep=5pt,minimum size=5pt] (q_0) {\small $1$};
    \node[state,inner sep=5pt,minimum size=5pt] (q_1) [right=of q_0] {\small $2$};
    \node[state,inner sep=5pt,minimum size=5pt] (q_3) [right=of q_1] {\small $3$};

    \path[->]
    (q_0) edge [] node {$\lambda_{1}$} (q_1)
    (q_1) edge [] node {$\lambda_{2}$} (q_3);
\end{tikzpicture}
    \caption{Structure of $Y(t)$ describing the precipitation pattern, with $\lambda_1 = \lambda_2 = 1/15$.}
    \label{fig:ExampleY}
\end{figure}
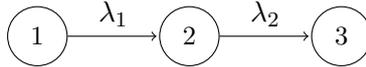

Besides incidents,
there may be other traffic events that
have a severe impact on the speeds the vehicle can drive.
For example, consider the situation that between $t\!=\!15$ and $t\!=\!30$
precipitation is forecasted. 
Note that, typically, 
precipitation does not
just affect the speeds around one link,
but has impact on a larger area of the network.
Now, to include these weather conditions in the background process $B(t)$,
we simply extend $B(t)$ by an additional Markov process $Y(t)$ that
describes whether at time $t$ the precipitation has not yet started~(encoded by $Y(t)\!=\!1$),
currently falls~($Y(t)\!=\!2$) or has already
stopped~($Y(t)\!=\!3$).
The three states are visited successively (Figure~\ref{fig:ExampleY}), 
with $Y(t)\!=\!1$ at $t\!=\!0$.
The extension of $B(t)$ allows us to define the speed levels on the arcs,
again, by letting the velocity on an arc $i$ equal $v_i(s)$
whenever $B(t)\!=\!s$. We have thus constructed the background process 
$B(t) := (X_1(t),\dots,X_6(t),Y(t))$ with $2^6\cdot 3$ states.

\begin{remark}\label{rem:R1}
{\em {In the above example, we model,
for simplicity, the durations of the first two 
states of $Y(t)$ by exponential distributions. However,
as we will argue in Section~\ref{subsec:fittingevents}, 
the {\sc mvm} framework is not restrictive, in the sense that we can
work with a considerably more general class of distributions. 
Importantly, these {\it phase-type distributions} can accommodate random quantities that are 
both less and more variable than the exponential distribution, thus making the setup highly flexible.}} $\hfill\Diamond$
\end{remark}

Notably, the precipitation may not only 
impact the driveable
speed levels on the arcs, but may, additionally,
impact the incident rate on the arcs 
\citep{Bergel-Hayat2013ExplainingEffects}.
Therefore, we allow the transition rates 
of the processes $X_i(t)$
to depend on the state of $Y(t)$.
Then, with $Q_y$ the transition rate matrix
of $(X_1(t),\dots,X_6(t))$ in case $Y(t)=y\in\{1,2,3\}$,
the $(2^6\cdot 3)\times (2^6\cdot 3)$-dimensional transition rate matrix $Q$ of the
background process $B(t)$ is of the form
\begin{align*}
    Q = \begin{bmatrix}
    Q_1\!-\!\lambda_1I & \lambda_1 I & \\
    & Q_2\!-\!\lambda_2I & \lambda_2I \\
    & & Q_3
    \end{bmatrix},
\end{align*}
with $\lambda_1$ and $\lambda_2$ as in Figure \ref{fig:ExampleY}.

\begin{wrapfigure}{r}{.35\textwidth}
    \includegraphics[width=\linewidth]{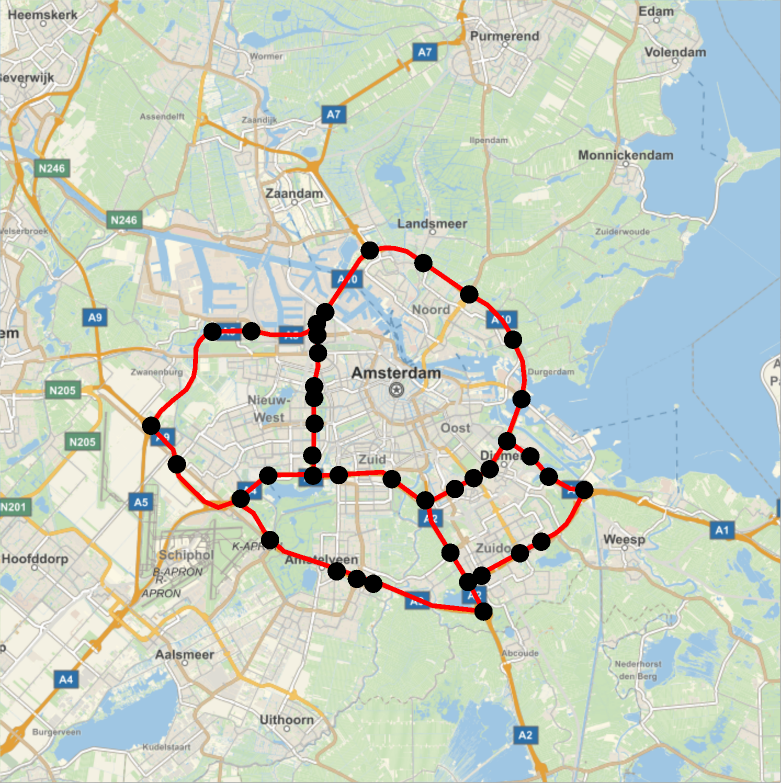}
    \caption{Amsterdam highway network, with nodes representing
    ramps, and links representing
    the two directed arcs between these ramps.}
    \label{fig:ringAmsterdam}
\end{wrapfigure}
In case traffic incidents and the upcoming precipitation are the only
events potentially affecting arc speeds, 
the full travel time distribution of the considered 
vehicle can be derived
from the presented random speed dynamics.
Indeed, given we know the state of the background process upon departure from Almelo,
the above formalism specifies the distribution of the time it takes to arrive in Hengelo.
Evidently, if there are additional sources of variability, 
the background
process can be extended to include these
events as well. 
Details on the general structure
of $B(t)$ are given in the mathematical
model description below.

\subsection{Mathematical model description} \label{subsec:mathmodel}~\\
After having introduced various concepts in our illustrative example, 
let us now consider a general road network, encoded by 
its corresponding graph representation $G = (N,A)$, 
of which the set of nodes $N$ represents the ramps
in the road network, and the set of directed arcs $A$ represents the roads
connecting these ramps. 
Hence, $k\ell \in A$ only if the ramps
represented by the nodes $k$ and $\ell$ are subsequent
ramps on one (directed) road.
A small example is the path-network between Almelo-Hengelo,
as shown in Figure~\ref{fig:AH}.
An example network of more realistic size, the
highway network around Amsterdam, the Netherlands,
is displayed in
Figure~\ref{fig:ringAmsterdam}.
Note that splitting highways at the ramps allows the model
to be used for practical applications such as the routing
of individual vehicles.
In the rest of this paper, both the terms \textit{arc}
and \textit{link} refer to a
highway part that arises by this splitting procedure,
i.e., a piece of highway enclosed by two ramps.
In case we consider a highway part enclosed
by two highway intersections, we will use
the term highway \textit{segment}. Note that,
with every intersection being a ramp but not vice versa,
a highway segment can always be partitioned into highway links.

In reality, the driveable speed level
on the link $k\ell \in A$
is not necessarily constant.
As argued, events such as incidents and heavy rainfall 
lead to fluctuations in the speeds vehicles can drive at.
As illustrated in the modeling example above,
the {\sc mvm} captures this randomness
in vehicle speeds by the introduction of an
environmental background process~$B(t)$ on the arcs~$A$,
which keeps track of the events affecting the arc speeds.
To this end, the {\sc mvm} distinguishes three types of events:
(i)~recurrent events, (ii)~random incidents, and (iii)~the 
{\mbox {(semi-)predictable}} non-recurrent traffic events that
are either present at the vehicle's departure
or known to occur in the foreseeable future
(e.g.\ forecasted snowfall, road work).
We will refer to the third type of events as
\textit{scheduled events}.

Let $n := |A|$, and write $A = \{a_1,\dots,a_n\}$ for the set
of arcs in $G$, with $a_i := k_i\ell_i$ for some $k_i,\ell_i \in N$.
To model traffic incidents, our primary focus,
we define $\{X_{a_i}(t),t\geqslant 0\}$ 
as independent Markov processes such that
for $a_i \in A$:
\begin{align*}
    X_{a_i}(t) &= 
    \begin{cases}
    1 &\quad \quad \text{if there is an incident on arc $a_i$ at time $t$,} \\
    2 &\quad \quad \text{otherwise}.
    \end{cases}
\end{align*}
Then, the Markovian background process $B(t)$ recording the `incident status' of the full
network $G$ is given by $(X_{a_1}(t),\dots,X_{a_n}(t))$.
We let the velocity of a vehicle traversing $a_i$
be determined by the background process $B(t)$
in the following way: 
if $B(t)$ is in state
$s \in \{1,2\}^n$,
the speed at which vehicles are moving on the arc
is $v_{a_i}(s)$.
This way, the speed on arc $a_i$
is allowed to depend on 
all processes $X_{a_1}(t),\dots,X_{a_n}(t)$.
Notably, the possibility to model 
correlation between
speeds on different arcs 
is an important asset of the {\sc mvm}, as
it can be used to model real-world traffic
phenomena like the spillback effect.

Note that the current structure of the processes $X_{a_i}(t)$ is such
that there are only two states for the modeling of incidents
and their impact on arc speeds. We are, however,
by no means restricted to this two-state structure: we could allow $X_{a_i}(t)$
to be any continuous-time Markov process. This gives us the opportunity
to model more complex incident speed patterns (e.g.\ distinguishing the incident itself, a recovery phase, and 
the regular conditions), and additionally provides
flexibility for the distribution of the incident length (i.e., this distribution is 
no longer strictly exponential; see Remark \ref{rem:R1}). 
Thus, we let $X_{a_i}(t)$ be
a continuous-time Markov process that represents
the state of an incident at arc $a_i \in A$ at time $t$,
and set $B(t) = (X_{a_1}(t),\dots,X_{a_n}(t))$,
with $X_{a_i}(t), X_{a_j}(t)$ evolving 
independently for $i\!~\neq\!~j$.
Dependence between the arc speeds is again realized by 
allowing the velocity on each arc to depend
on the state of the vector $B(t)$:
if $B(t) = s$ the velocity
at which vehicles are moving on arc $a_j$ is
$v_{a_j}(s)$.

To include the two other types of events, the recurrent
and the scheduled events, we expand
the background process $B(t)$ with a Markov process $Y(t)$.
Specifically, in case there are $m$ scheduled events, $Y(t)$ is
structured as $(Y_0(t),Y_1(t),\dots,Y_m(t))$,
with $Y_0(t)$ a Markov process that models the
effect of the recurrent, daily traffic patterns, 
and $Y_1(t),\dots,Y_m(t)$ Markov processes that model
the $m$ scheduled events (of which
the process in Figure~\ref{fig:ExampleY} is an example). 
As noted in the illustrative
example of Section~\ref{subsec:modelingexample}, 
recurrent and scheduled events may not just impact
the driveable speeds in the network, but may,
additionally, impact the transition rates of 
the processes $X_{a_i}(t)$.
For example, for an arc in the network,
there may be a significant difference between
the inter-incident time within and outside
the rush hours. Therefore, we 
let $B(t) = (Y(t),X_{a_1}(t),\dots,X_{a_n}(t))$
be such that only conditional on the state of 
the `common process' $Y(t)$, the individual processes
$X_{a_i}(t)$ (for $i=1,\ldots,n$) evolve independently.

Now, in case the background process $B(t)$ for a departing
vehicle is fully specified, i.e., all background states and
transition rates are known, the travel time distribution
is fully specified as well. That is, if $B(t)$ is in state $s$ at
the departure time of the vehicle, the travel time
on an edge $a_i$ with length
$d_{a_i}$ is distributed as $\tau^s_{a_i}$,
with
\[\tau_{a_i}^{s}:= \min\left\{t\geqslant 0: \left.
\int_{0}^{t} v_{a_i}( B(u))\,{\rm d}u
\geqslant d_{a_i} \:\right|\, B(0) = s\right\}.\]
An expression for the Laplace-Stieltjes transform~(LST) of 
$\tau^s_{a_i}$ was derived \cite{Levering2022AConditions}.
Importantly, the LST of a non-negative random variable uniquely 
determines its distribution function, and, moreover, the derivatives of
the LST yield the moments of the random variable.

\subsection{Fitting traffic events} \label{subsec:fittingevents}~\\
One of the advantages of the {\sc mvm} framework is its flexibility, in the sense that
it allows a high degree of generality when it comes to the distributions of the durations of the underlying events. 
It is true that the times spent in the states of a continuous-time Markov process
necessarily follow exponential distributions, but, as briefly mentioned in Remark \ref{rem:R1}, the {\sc mvm} is still capable of handling non-exponential distributions.
That is, if data analysis would reveal either the duration of an event 
affecting arc speeds, or the duration between two such events, to be non-exponential,
we can use {\it phase-type distributions} to cast these durations into the Markovian
setting, and, as a result, include the event(s) into the {\sc mvm}.
Importantly, phase-type distributions have the attractive property that they can model random quantities 
that are less variable than the exponential distribution 
as well as random quantities that are more variable 
than the exponential distribution. In the remainder of this subsection we provide more background.

Informally, the class of phase-type (PH) random variables consists of all sums and mixtures of exponentially
distributed random variables.
This concretely means that any phase-type distribution is characterized 
by a Markov process with $d\!+\!1$ states,
an entrance probability vector $\boldsymbol{\alpha} \in \mathbb{R}^{d+1}$,
and a transition rate matrix of the form
\begin{align*}
Q_{\rm PH} = 
\begin{pmatrix}
T & -T\boldsymbol{1} \\
\boldsymbol{0}^\top & 0\\
\end{pmatrix},
\end{align*}
with $T \in \mathbb{R}^{d \times d}$, and $\boldsymbol{0}$ and $\boldsymbol{1}$ respectively denoting all zeroes and all ones
$d$-dimensional column vectors; see \cite[Section III.4]{Asmussen2003AppliedQueues}.
The transient states $1,\dots,d$ are the so-called {\it phases} of the Markov process.
From the structure of the transition matrix $Q_{\rm PH}$, we note that state $d\!+\!1$ is an absorbing state.
With the random variable $X$ denoting the total elapsed time from the start
of the described Markov process 
until absorption in $d\!+\!1$, we say that $X$ has a PH($\boldsymbol{\alpha},Q_{\rm PH}$)
distribution.
From this definition, it is immediately clear that we can include
any traffic event with a PH duration into our framework.
Instead of working with a single phase, as we did in the above examples with exponentially distributed durations, 
we now include the $d$ 
phases of the PH distribution into our model. 
Concretely, when the event starts, the initial phase is sampled according to $\boldsymbol{\alpha}$, after which
the Markov process $X$ evolves according to $Q_{\rm PH}$
until a transition to the absorption state occurs. 
Then the background process of the {\sc mvm} moves to one
of the outdegree neighbors belonging to the event under consideration,
according to their respective transition rates.

We already mentioned that phase-type distributions can model non-negative random quantities 
that differ in variability from the exponential distribution. We proceed by making this claim more precise. 
To obtain an approximating distribution for such a random quantity $X$,
it is common procedure to use the {\it two-moment phase-type
matching approximation} that was advocated by
\cite{Tijms1986StochasticApproach}.
The underlying principle is to fit a phase-type distribution
to the mean $\mathbb{E}[X]$ and the {\it squared coefficient of variation} $c_X^2$,
defined as:
\begin{align*}
c_X^2 &= \frac{\text{Var}(X)}{\mathbb{E}[X]^2} = \frac{\mathbb{E}[X^2]}{\mathbb{E}[X]^2}-1.
\end{align*}
Note that the SCV of an exponentially
distributed random variable equals 1.
The fitting procedure distinguishes two cases:
\begin{itemize}
\item[$\circ$] In case $0\!<\!c_X^2\!<\!1$, the distribution of $X$ is less variable than the exponential distribution.
In this case $X$ is approximated by
a mixture of Erlang distributions.
That is, the fitted distribution is 
with probability~$p$ an Erlang distribution with $k\!-\!1$ phases and
mean $(k\!-\!1)/\mu$, and with probability~$1\!-\!p$
an Erlang distribution with $k$ phases and mean $k/\mu$.
A simple calculation shows that 
the SCV of this distribution equals $(k-p^2)/(k-p)^2$,
which for $p\in [0,1]$ lies between $1/k$ and
$1/(k\!-\!1)$. Hence, we set~$k$ 
such that $1/k\!\leq\!c_X^2\!\leq\!1/(k\!-\!1)$. 
Both $\mu$ and $p$ are now chosen such that 
the mean and the SCV of the mixture Erlang distribution 
uniquely match $\mathbb{E}[X]$ and $c_X^2$.
\item[$\circ$] In case $c_X^2\!\geq 1$, the distribution of $X$ is more variable than the exponential distribution.
In this case $X$ is approximated by
a hyperexponential distribution, which equals an exponential($\mu_1$) distribution
with probability $p$, and an exponential($\mu_2$) distribution with probability $1\!-\!p$.
In this setting, the three parameters cannot be uniquely determined from
$\mathbb{E}[X]$ and $c_X^2$. However, this can be tackled 
by imposing \textit{balanced means}, i.e., 
using the normalization $p/\mu_1 = (1\!-\!p)/\mu_2$, which reduces
the number of free parameters from three to two.
\end{itemize}

\begin{figure}[ht]
\centering
  \begin{subfigure}[b]{0.35\textwidth}
    \centering
    \begin{tikzpicture}[shorten >=1pt,node distance=1cm,on grid,auto]
    \node[state,inner sep=5pt,minimum size=5pt] (1) {};
    \node[draw=none,inner sep=5pt,minimum size=5pt] (non1) [right=1.5 of 1] {};
    \node[state, inner sep=5pt,minimum size=5pt] (2) [right=2 of 1] {};
    \node[draw=none,inner sep=5pt,minimum size=5pt] (non2) [right=of 2] {};
    \node[state,inner sep=5pt,minimum size=5pt] (3) [right=of non2] {};
    \node[state,white,inner sep=5pt,minimum size=5pt] (1a1) [above=0.5 of non1] {};
    \node[state,white,inner sep=5pt,minimum size=5pt] (1a2) [below=0.5 of non1] {};
    \node[draw=none,inner sep=5pt,minimum size=5pt] (non2a) [right=0.5 of non1] {};
    \node[draw=none,inner sep=5pt,minimum size=5pt] (non3a) [right=0.5 of non2a] {};
    \node[state,white,inner sep=5pt,minimum size=5pt] (3a2) [below=0.5 of non3a] {};

    \path[->]
    (1) edge [] node {} (2)
    (2) edge [] node {} (3);
\end{tikzpicture}
    \caption{Exponential precipitation duration.}
    \label{fig:fittingPHa}
  \end{subfigure}
  \quad \quad \quad
  \begin{subfigure}[b]{0.35\textwidth}
    \centering
    \begin{tikzpicture}[shorten >=1pt,node distance=1cm,on grid,auto]
    \node[state,inner sep=5pt,minimum size=5pt] (1a) {};
    \node[draw=none,inner sep=5pt,minimum size=5pt] (non1a) [right=1.5 of 1a] {};
    \node[state,inner sep=5pt,minimum size=5pt] (1a1) [above=0.5 of non1a] {};
    \node[state,inner sep=5pt,minimum size=5pt] (1a2) [below=0.5 of non1a] {};
    \node[draw=none,inner sep=5pt,minimum size=5pt] (non2a) [right=0.5 of non1a] {};
    \node[draw=none,inner sep=5pt,minimum size=5pt] (non3a) [right=0.5 of non2a] {};
    \node[state,inner sep=5pt,minimum size=5pt] (3a2) [below=0.5 of non3a] {};
    \node[state,inner sep=5pt,minimum size=5pt] (4a) [right=1.5 of non3a] {};

    \path[->]
    (1a) edge [] node {} (1a1)
    (1a) edge [] node {} (1a2)
    (1a1) edge [] node {} (4a)
    (1a2) edge [] node {} (3a2)
    (3a2) edge [] node {} (4a);
\end{tikzpicture}
    \caption{Mixture Erlang precipitation duration.}
    \label{fig:fittingPHb}
  \end{subfigure}
    \caption{Modeling the precipitation length,
    instead of with an exponential distribution (a), with
    a mixture Erlang distribution (b).}
  \label{fig:fittingPH}
\end{figure}
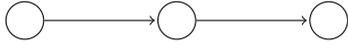
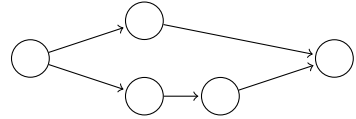

\begin{wrapfigure}{r}{.28\textwidth}
    \vspace{-17pt}
    \begin{minipage}{\linewidth}
    \centering\captionsetup[subfigure]{justification=centering}
    \includegraphics[width=\linewidth]{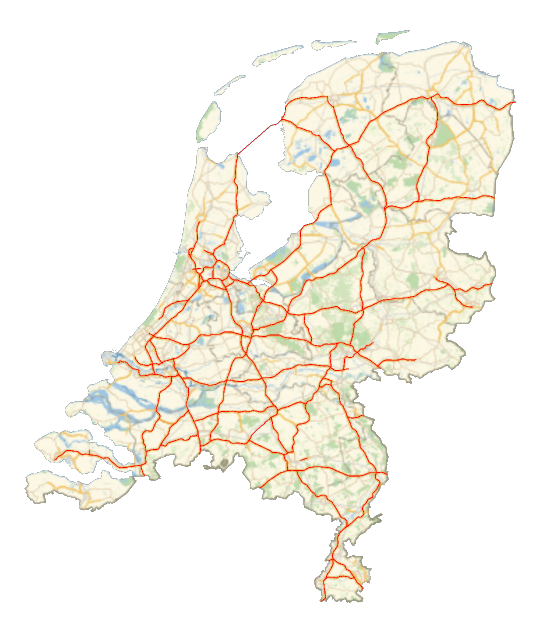}
    \par\vfill
    \includegraphics[width=\linewidth]{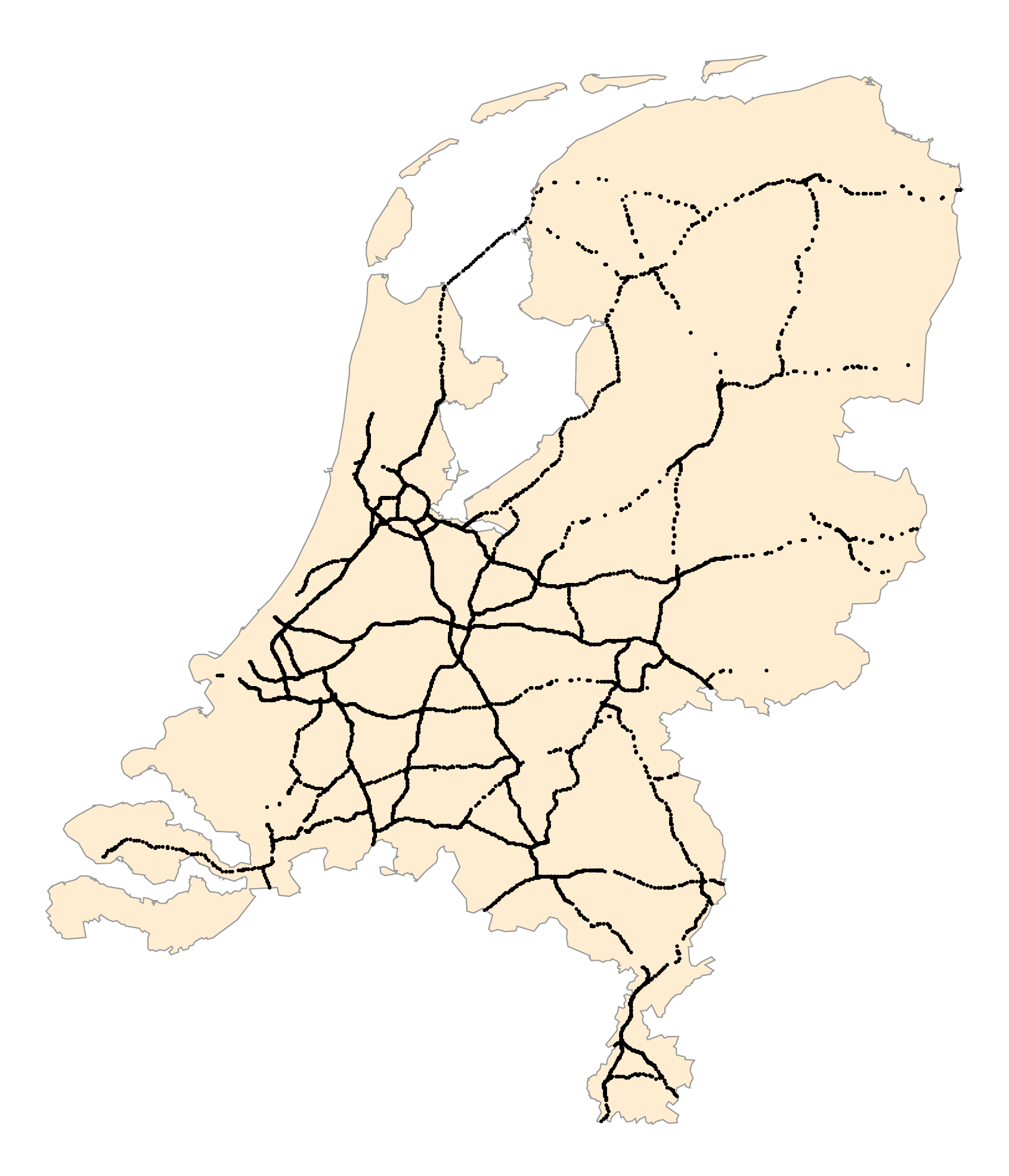}
\end{minipage}
    \caption{The segments (above) and loop detectors (below)
    in the Dutch highway system.}
    \vspace{-5pt}
  \label{fig:dutchnetwork}
\end{wrapfigure}
A small example for the implementation of phase-type distributions
in $B(t)$ is provided in Figure~\ref{fig:fittingPH}.
We consider the forecasted precipitation from 
Section~\ref{subsec:modelingexample},
whose impact was described by a Markov
process with successive states 1, 2 and 3,
respectively denoting the time before, during and after
the precipitation (Figure~\ref{fig:fittingPH}a).
Consequently, both the time until the precipitation
and the duration of the precipitation are
modeled by exponential distributions.
However, if statistical analysis and the above fitting procedure reveal
that the duration of
the precipitation is better described by
a mixture of an exponential distribution
and the sum of two exponential distributions, 
we can include this PH distribution
by replacing state 2 by the three phases
of this distribution (Figure~\ref{fig:fittingPH}b).

\section{Speed pattern analysis in the Dutch highway network} \label{sec:dataanalysis}
This section investigates daily
speed patterns, and, in particular, the
impact of traffic incidents
on these driveable vehicle speeds.
The study focuses on the Dutch highway network, 
for which extensive data sets on traffic jams
and vehicle speeds are openly
available (Section~\ref{subsec:networkanddata}).
Importantly, use of the Dutch data is merely illustrative,
as the techniques we present can be
applied to any highway network for which
similar data sets are accessible.
Now, to describe the traffic patterns in the Dutch 
network, we investigate, for all individual segments, 
the random durations
and attained velocities in both the inter-incident
(Section~\ref{subsec:interincident}) and
the incident regimes (Section~\ref{subsec:incidents}).
It turns out that these durations and
corresponding speed levels depend on the time of day
and day of week, but that there are periods in which
these effects are relatively constant.

In principle, the insights of this section can be used 
in any model describing vehicle speeds.
The {\sc mvm} of Section~\ref{sec:MVM} is an example of such
a model, and we will argue that the findings
of the present section indeed align with the {\sc mvm}.
To this end, we already include some remarks in the present
section that reveal how the observed
speed patterns can be incorporated in the {\sc mvm}.
More detail is provided in Section~\ref{sec:oper}, in which we
further specify the fitting procedure, and highlight how to
construct the background process
of the {\sc mvm} corresponding to the Dutch highway
network.

\subsection{Network and data description} \label{subsec:networkanddata}~\\
The Dutch highway network, depicted
in Figure~\ref{fig:dutchnetwork}, consists
of all highways (i.e.,\ the so-called A-roads) in the Netherlands.
In our analysis, we additionally include three non-highway
trajectories, that each serve as important connection
between two highways. Now, to study the speed levels
in this network, we split each of these roads into highway segments,
i.e., highway parts separated by highway interchanges.
Hence, each network segment we consider 
is either a directed road between two
highway interchanges, or the first or last
section of a highway.
Splitting at the interchanges
allows the use of the results
for the travel time estimation
of vehicles traversing paths
consisting of multiple highways.
Moreover, splitting a highway into a
larger number of segments implies that 
an even larger amount of data
will be required for a meaningful
analysis 
of the incident impact.

\begin{wrapfigure}{r}{.35\textwidth}
    \begin{minipage}{\linewidth}
    \centering\captionsetup[subfigure]{justification=centering}
    \includegraphics[width=\linewidth]{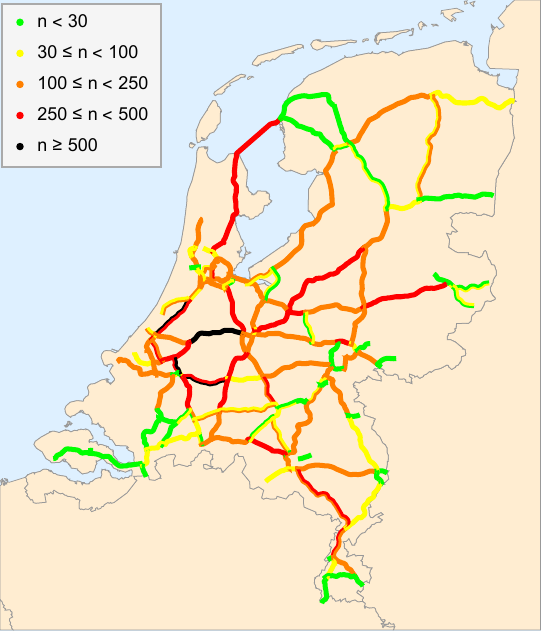}
    \subcaption{Location}
    \label{fig:interincloc}
    \par\vfill
    \includegraphics[width=\linewidth]{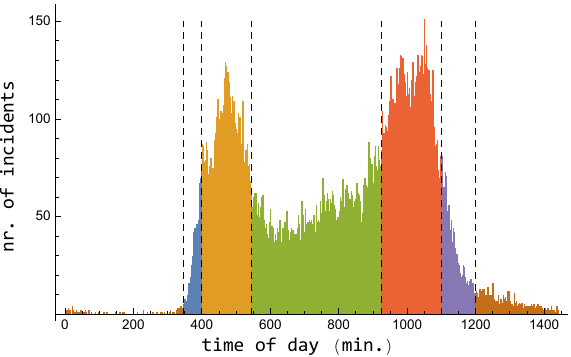}
    \subcaption{Time of day}
    \label{fig:interinctime}
    \end{minipage}
    \caption{Dependence of number of registered incidents ($n$)
    in the years 2015-2019 on location and time of the day.}%
    \label{fig:dependence}%
\end{wrapfigure}
We study traffic data of the Dutch highway network
to predict the impact of traffic events on the velocities
at the highway segments.
Specifically, we assess the incident duration,  
the time between incidents,
and the vehicle speeds in both these settings.
For the analysis, two openly available data sets are used: 
(i) a database with traffic speeds and flows 
at loop detectors in the Dutch road network and 
(ii) a list of registered traffic jams.
In the Netherlands, these two data sources
are managed by the National Road Traffic Data Portal ({\sc ndw})
and Rijkswaterstaat~({\sc rws}), respectively.

The {\sc ndw} data set \citep{ndwdatabase} 
contains loop detector data from the year 2017.
On most Dutch highways, there is a high concentration of
loop detectors, as can be noted from their locations 
as shown in Figure~\ref{fig:dutchnetwork}. Every minute,
the average traffic flow and speed at these loops 
are registered and stored.

The {\sc rws} data set 
\citep{rwsdatabase}
contains all registered traffic jams 
from the year 2015 onward,
of which we use the registrations from the years 2015-2019.
Even though data for the years 2020-2021 is available as well,
we exclude these years from our analysis, due to a change in traffic conditions.
First, early in 2020, the Dutch government introduced reduced
speed limits during the day.
Second, during the Covid-19 pandemic of the years 2020-2021,
the Dutch government imposed a
\textit{working-from-home} measure which led to significant 
reductions in traffic flows.
Note that the {\sc rws} files contain {\it all} 
registered traffic jams, whereas for our
study on the impact of incidents we 
only used the entries for which the cause of
the traffic jam is marked as `incidental' (i.e., caused by an incident).
Examples of non-incidental causes 
include rush hours and planned road works.
After cleaning of the incident data, 
the database contains 58152 incident
entries.

\subsection{Time between incidents} \label{subsec:interincident}~\\
We start the analysis by considering traffic 
in non-incident state.
In this setting, there are two main objectives: (i)~fit a distribution
on the length of this state, i.e., the time between incidents,
and (ii)~estimate the corresponding vehicle speed level.
For these objectives, it is important to note, 
as will be shown below, that both the
time between incidents and the vehicle speed levels are location- and time-dependent.
We deal with the location-dependence by considering the inter-incident time per
highway segment. We deal with the time-dependence by identifying periods within
which time has hardly any impact on the inter-incident duration and the speed level.

\begin{table}[ht]
\begin{center}
\begin{tabular}{ |c||c|c|c|c|c|c|c|} 
 \hline
  & Min & Q1 & Q2 & Q3 & Max & Mean & St.dev. \\ \hline
 Incident duration (in min.) & 0.1 & 23.4 & 43.2 & 71.3 & 1041.0 & 54.9 & 48.6 \\ 
 Inter-incident time (in min.) & 1.0 & 1460.5 & 5209.4 & 11901.7 & $1.4 \cdot 10^6$ & 12279.0 & 33714.7 \\ 
 \hline
\end{tabular}
\end{center}
\caption{Scale-comparison incident duration and inter-incident duration.}
\label{tab:durationvsbetween}
\end{table}

In the first place, as observed from Figure~\ref{fig:dependence}a,
the frequency of incidents differs throughout the highway network.
Therefore, we estimate the inter-incident time per highway segment.
Then, as observed from Table~\ref{tab:durationvsbetween}, 
the durations of incidents are short relative to the time between incidents.
Thus, we may redirect our focus, and consider, 
instead of the time between incidents,
the time between the {\it start} of two consecutive incidents.
Satisfying the memoryless property, 
the exponential distribution is widely used
to model the time between two elapsed events.
Now, if the time between two incident starts
would indeed fit an exponential distribution, 
the occurrence of incidents could be modeled
by a homogeneous Poisson process, and, consequently, the starting
time of incidents would be uniformly distributed over the time of day.
However, the incident starts in Figure~\ref{fig:dependence}b 
do not show a uniform pattern.
Therefore, the time between two incidents is unlikely to 
follow an exponential distribution.

\begin{wrapfigure}{r}{0.35\textwidth}
  \begin{center}
    \includegraphics[width=0.95\textwidth]{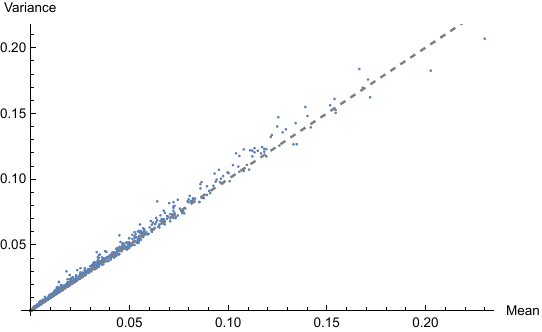}
  \end{center}
\caption{Total number of registered incidents: mean versus variance.}
	\label{fig:meanvsvar}
\end{wrapfigure}
Nevertheless, we observe that there are periods in Figure~\ref{fig:dependence}b
in which the number of incidents is approximately
uniform. Within these periods, the exponential distribution
{\it would} be a promising fit for the time between 
two consecutive incident starts.
To check if a partition into periods could in fact offer a solution to the observed time-dependence,
we identify six periods, as separated
by the dashed lines in Figure~\ref{fig:dependence}b, in which the 
number of incidents is roughly uniformly
distributed.
Modeling
the time between two consecutive incident starts
on a segment in a single period
with an Exponential($\lambda$)-distribution
corresponds to a Poisson($\lambda t^*$)-distribution for the number of incidents in that period, 
with $t^*$ denoting the period length.
Indeed, from Figure~\ref{fig:meanvsvar}, which plots, per segment,
the mean numbers of registered incidents in an elapsed period against the corresponding variances,
we find that the points are concentrated around the $y=x$ line, which is indicative of the Poisson distribution providing a good fit.

\begin{figure}[b]
  \begin{center}
    \includegraphics[width=0.9\textwidth]{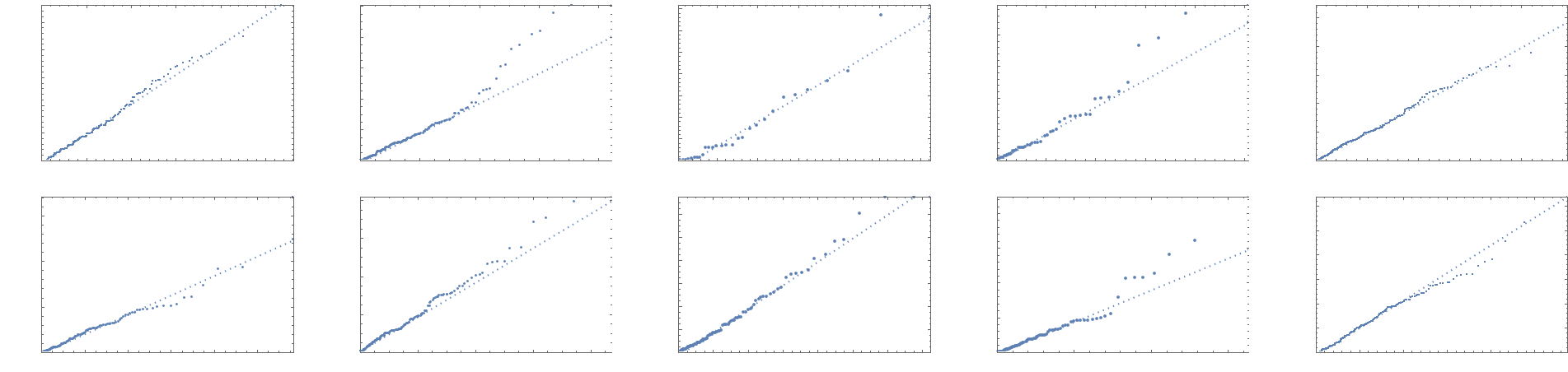}
  \end{center}
\caption{QQ-plots for simulated time between incidents (horizontal axis) against real-world observations (vertical axis). With inter-incident
times of high-order, the axes values
are omitted for display purposes.}
	\label{fig:qqtimebetween}
\end{figure}

To further assess the quality of the exponential fits over the six
periods, we compare simulation results with the {\sc rws} data.
We plot, for ten randomly selected segments,
the quantiles of the empirical data distribution of 
the time between incidents against quantiles of 
simulations of the time between
the start of two consecutive incidents (Figure~\ref{fig:qqtimebetween}).
All QQ-plots show a high degree of linearity between the quantiles, 
thus corroborating the exponentiality claim.
There are some small deviations for large inter-accident times, but we note
that these will have little effect on travel time predictions,
as travel times generally relate to a considerably smaller timescale.

Two additional remarks on the presented fitting procedure:
\begin{itemize}
\item[$\circ$] We have identified six periods 
in which the number of registered incidents
behaves roughly uniformly. An even better fit could 
potentially be obtained by dividing the time-frame into
more periods. However, increasing
the number of periods will decrease the number
of observations per (segment, period)-pair,
and may therefore lead to less reliable
results. 
Moreover, 
a favorable consequence of working with a relatively
low number of periods is the 
resulting
low complexity of the resulting {\sc mvm}.
\item[$\circ$] Period transitions have been 
chosen to occur simultaneously at every
segment of the network. 
A potential better fit could
be found by adding more detail with
a period division per segment, or subset of segments.
However, again, the number of observations
per segment may limit the quality of these
more detailed partitions.
Note that the uniform choice
in period transitions over all segments
also has an advantage when fitting the {\sc mvm}:
we are able to include the duration of these periods
in $Y(t)$ (instead of including them
in $X_{a_i}(t)$ for all $i~=~1,\dots,n$),
which will keep the computational complexity low.
\end{itemize}

\begin{figure}[ht]
\centering
  \begin{subfigure}[b]{0.3\textwidth}
    \centering
    \includegraphics[width=\textwidth]{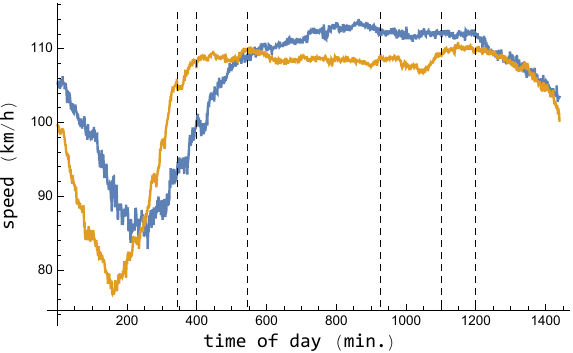}
    \caption{Part of A67 highway.}
    \label{fig:AvMaxSpeedsA67}
  \end{subfigure}
  \quad
  \begin{subfigure}[b]{0.3\textwidth}
    \centering
    \includegraphics[width=\textwidth]{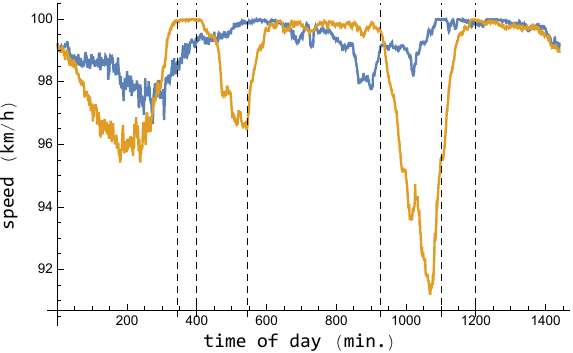}
    \caption{Part of A10 highway}
    \label{fig:AvMaxSpeedsA10}
  \end{subfigure}
    \caption{Average maximum speeds 
at weekends (blue) 
and weekdays (yellow) in 2017. Vertical
lines partition the periods
characterized in Section~\ref{subsec:interincident}.}
  \label{fig:AvMaxSpeeds}
\end{figure}

We have established that the time between two incidents 
in each of the periods in Figure~\ref{fig:dependence}b 
can be modeled by an exponential distribution.
For a description of the traffic patterns 
in this non-incident state, 
we study the corresponding vehicle speeds.
To this end, we have available loop detector data
on traffic speeds and traffic flows, provided by 
{\sc ndw}. For a given segment, we collect, 
per minute and per detector located at this segment,
the maximum over the registered average 
speeds of the road lanes.
Note that by taking the maximum of the averages we
limit the impact of slow-moving vehicles on
the collected speed levels. Indeed, we are interested in 
the per-segment {\it potential} driveable speed,
which is not well reflected by data corresponding to 
e.g.\ trucks. Since the driveable speed
cannot exceed the speed limit, we additionally upper bound
these maximum average speeds by the speed limit of
the corresponding segment.

\begin{wrapfigure}{r}{.3\textwidth}
    \includegraphics[width=0.95\linewidth]{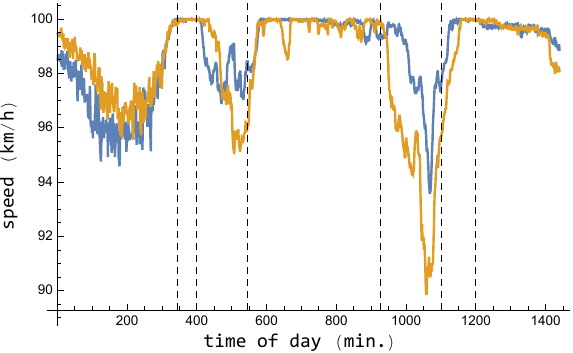}
    \caption{Average maximum speeds on the A10 trajectories at weekdays 
    in February (blue) and May (yellow) in 2017, with rush hour speeds
    significantly lower for the month May.}
    \label{fig:AvMaxSpeedsSeasonality}
\end{wrapfigure}
For expositional reasons, we present
the results of the loop detector data
study for two representative highway
trajectories: a part of the busy A10 highway
around the city of Amsterdam, and a part of the 
less traveled A67 highway in the southern
part of the Netherlands.
Figures~\ref{fig:AvMaxSpeedsA67} and~\ref{fig:AvMaxSpeedsA10}
show the average maximum speeds 
per minute of the day for 
the A67 and A10 trajectory respectively.
From these plots it can be observed that,
similar to the duration of 
the inter-incident state,
vehicle speeds are time-dependent.
This is most notable in Figure~\ref{fig:AvMaxSpeedsA10},
in which the speed patterns around the rush hours
clearly differ from the patterns outside the rush hours.
This time dependence cannot
purely be explained by the effect of the time of day,
as e.g.\ the speed patterns during days in the weekend
are significantly different from the speed patterns during
weekdays. Moreover, besides time of day and day of week,
we observe that there are also seasonal effects (Figure~\ref{fig:AvMaxSpeedsSeasonality}).
Evidently, when estimating travel time distributions, these
temporal influences should be taken into account.

\begin{figure}[ht]
\centering
  \begin{subfigure}[b]{0.3\textwidth}
    \centering
    \includegraphics[width=\textwidth]{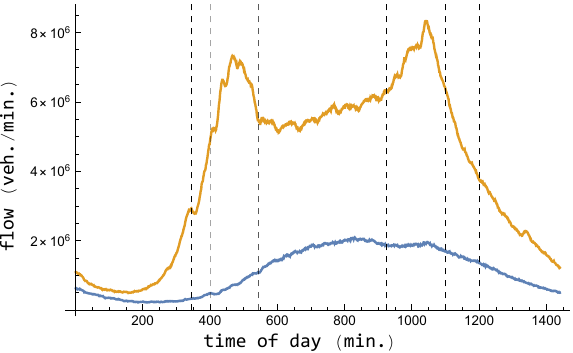}
    \caption{Part of A67 highway.}
    \label{fig:flowA67}
  \end{subfigure}
  \quad
  \begin{subfigure}[b]{0.3\textwidth}
    \centering
    \includegraphics[width=\textwidth]{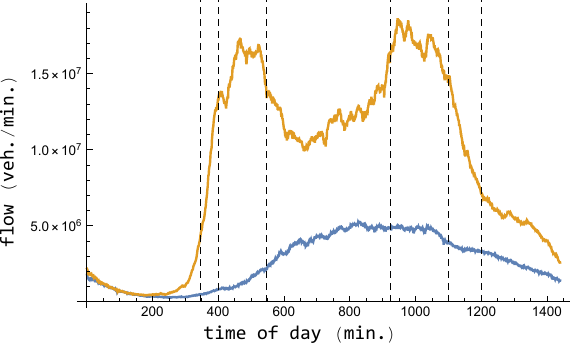}
    \caption{Part of A10 highway.}
    \label{fig:flowA10}
  \end{subfigure}
    \caption{Average traffic flows 
on parts of the Dutch
highway network at weekends (blue) 
and weekdays (yellow) in 2017. Vertical
lines partition the periods
characterized in Section~\ref{subsec:interincident}.}
  \label{fig:AvMaxFlows}
\end{figure}

When considering
the periods as characterized in
Figure~\ref{fig:dependence}b,
there is one period
for which the
effects of the day of the week
and month of the year 
should play an insignificant role: the night
period (8:00pm--6:45am).
Since traffic demands in this time interval
are typically low,
the free-flow speed should always 
be a good proxy 
for the mean car speed in non-incident state.
However, 
Figure~\ref{fig:AvMaxSpeeds}
shows that the attained speed levels during the night are typically low;
note that with a relatively high percentage
of slow-moving vehicles traveling at night,
the resulting imbalanced traffic mix is a probable explanation for this fact. 
To compare, the mid-day period (09:05am--03:25pm) generally
experiences
much higher flow levels (Figure~\ref{fig:AvMaxFlows}), but has speed levels
that \textit{are} close to the maximum speed. 
We therefore conclude
that in periods with much lower traffic flows (such as the nights), cars are able
to drive at those maximum speeds as well.

For most segments, speed levels outside
the night period are either affected by
the day of the week or the month of the year.
We will use the velocity data to deduce a few
conclusions about the speed patterns in
these periods, which will be useful
for the prediction of 
travel time distributions.
We focus on the A10 and A67 highway segments 
of Figure~\ref{fig:AvMaxSpeeds}, but similar
conclusions can be drawn for other highway segments:
\begin{itemize}
    \item[$\circ$] For both highway segments, 
    on weekend days, the highway
    speed limit is a representative speed level
    for all six periods.
    Similar to our reasoning above for the night period, 
    the daily flow levels
    on these days are typically too low to reduce the
    driveable vehicle speeds. 
    \item[$\circ$] Figure~\ref{fig:AvMaxSpeedsA67}
    shows that on weekdays, the A67 speed levels
    in the five non-night periods are fairly constant.
    Indeed, only the first morning rush hour period shows
    an increasing trend, but the corresponding low traffic flow
    exposes that, in this period, the actual driveable speeds
    are close to the speed limit.
    Therefore, in each of the periods, a
    constant speed level would
    serve as good representative 
    for the driveable speed. 
    Note, however, that the appropriate representative speed
    level may be dependent on the day of the week or 
    the month of the year.
    \item[$\circ$] Figure~\ref{fig:AvMaxSpeedsA10}
    displays that, on a non-weekend day, working with 
    one representative speed level
    for the A10 segment will certainly work well
    in the early morning and mid-day periods.
    The speed patterns in the other periods are not
    well described by a constant speed level, but do
    show a distinguishing pattern. That is, around the rush hours, 
    the high traffic flows
    clearly affect the driveable speeds, 
    showing an approximate V-shape
    for the speed drops. The period between the evening rush
    hour and the night period typically serves as recovery period,
    with decreasing flow levels and, consequently, increasing speed
    levels.
\end{itemize}
In conclusion, the speed pattern of a period is generally either well
represented by one constant speed level, or captured by (a
combination of) decreasing or increasing speed trends.
Working with the {\sc mvm} to model travel time distributions,
the speed levels belonging to the different
background states need to be chosen such that they 
emulate these speed patterns. 

\subsection{Incidents} \label{subsec:incidents}
\begin{wrapfigure}{l}{0.33\textwidth}
  \begin{center}
    \includegraphics[width=0.95\textwidth]{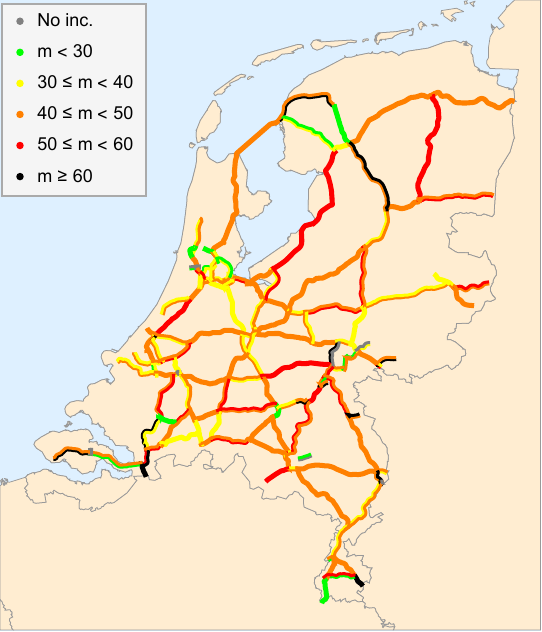}
  \end{center}
\caption{Median incident duration ($m$) per segment (in min.).}
	\label{fig:medianduration}
\end{wrapfigure}

~\\
We continue our analysis by considering incidents.
The objectives are in line with those of the inter-incident setting:
(i) fit a distribution on the incident duration,
and (ii) study the corresponding vehicle speeds.
Similar to the inter-incident time, we study
incidents per segment, as their duration depends
on their location (Figure~\ref{fig:medianduration}).
In the previous subsection, we 
characterized six periods in Figure \ref{fig:dependence}b
in which the number of incidents on a segment is roughly uniform.
We will show that, additionally,
these six periods suffice to deal with time-dependence
in incident length, and describe how to fit a distribution
for the incident duration for each (segment, period)-pair.

\begin{figure}[b]
\centering
  \begin{subfigure}[b]{0.3\textwidth}
    \centering
    \includegraphics[width=\textwidth]{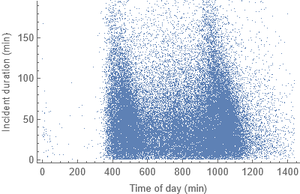}
    \caption{Raw data.}
    \label{fig:timedepincdurA}
  \end{subfigure}
  \quad
  \begin{subfigure}[b]{0.3\textwidth}
    \centering
    \includegraphics[width=\textwidth]{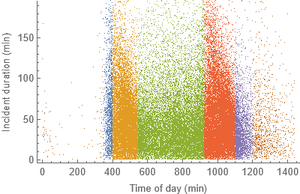}
    \caption{Split into six periods.}
    \label{fig:timedepincdurB}
  \end{subfigure}
   \begin{subfigure}[b]{0.3\textwidth}
    \centering
    \includegraphics[width=\textwidth]{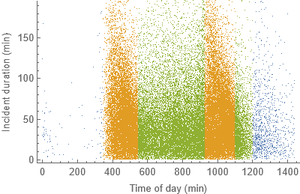}
    \caption{Split into three modes.}
    \label{fig:timedepincdurC}
  \end{subfigure}
\caption{Time of day of incident start (in min.) against length of the incident (in min.).}
\label{fig:timedepincdur}
\end{figure}

First, we observe from Figure~\ref{fig:timedepincdur}a
that incident length is indeed time-dependent.
More specifically, it can e.g.\ be seen that severe 
incidents (in terms of duration)
occur more frequently within rush hours than outside rush hours.
However, 
for the six periods identified for the inter-incident
time,
the correlation between 
the time of occurrence of incidents and their lengths 
is not significant,
as can be seen in Figure~\ref{fig:timedepincdur}b.
Thus, to describe the incident length on a highway segment, we can
fit a distribution for each
of these six periods.
We are, however, constraint by the amount of data 
per (segment, period)-pair.
Therefore, we further combine periods for which
the merged data is still approximately time-independent.
As a result, we obtain the three periods 
(yellow, green, and blue)
in Figure~\ref{fig:timedepincdur}c, and 
the objective is to fit a distribution to 
the incident duration
for each of these periods.

To find distributions for the incident lengths, 
we use the two-moment phase-type
matching approximation discussed
in Section~\ref{subsec:fittingevents}. We additionally fit
against the Erlang-1 (equivalent to exponential) and
Erlang-2 distribution, as these are phase-type distributions
with at most the complexity
of the hyperexponential and mixture Erlang distribution, 
making them even more preferable to work with. 
Initial fits show that outliers have a severe impact 
on the SCV, and consequently, a negative
impact on the fit. Thus, for every data set to fit, 
the 1$\%$ largest observations are excluded
in the computation of the SCV and estimation of the parameters.
Importantly, we \textit{do} include these outliers
in the data set when assessing
the quality of the obtained fits.

We start the fitting procedure by
considering the incident lengths in the
night period (Figure~\ref{fig:timedepincdur}c, blue).
Since, in this period, less than 5\% of the segments have more than 15 reported
incidents, we merge the observations and fit one distribution
that will be used to describe the incident length at night for all segments.
With the resulting SCV below~1, 
this approximating distribution is a mixture 
of Erlang distributions (see Section~\ref{subsec:fittingevents}).
For the two other periods, we fit an incident duration distribution 
per (segment, period)-pair, if there are  
sufficiently many data points for this pair.
In case the number of observations in one of the periods is low,
but the total number of observations on this segment in the two periods
{\it is} sufficient, we fit one distribution 
based on the joint set of observations.
For the remaining segments, we merge all observations in the rush hour 
period (Figure~\ref{fig:timedepincdur}c, yellow),
as well as in the non-rush hour 
period (Figure~\ref{fig:timedepincdur}c, green), and fit
a distribution on both these periods, used for all segments in this category.

\begin{table}[t]
\centering
\begin{tabular}{|c|c|c|c|c|c|} 
 \hline
\multirow{2}{*}{} & \multirow{2}{*}{$\%$ rejected} & \multicolumn{4}{c|}{$\%$ non rejected} \\
\cline{3-6}
& & Erlang-1 & Erlang-2 & Hyperexp. & Mixture Erl. \\ \hline
SCV $<1$ & $1.1\%$ & $1.1\%$ & $30.6\%$ & - & $68.3\%$ \\
SCV $\geq 1$ & $3.3\%$ & $68.3\%$ & $16.7\%$ & $15.0\%$ & - \\
 \hline
\end{tabular}
\caption{Fitting incident lengths for the (segment, period)-pairs. The last four columns show what
percentage of the non-rejected fits has the distribution of that column as highest p-value,
with p-values obtained through Anderson-Darling tests.}
\label{tab:incidentdurationfits}
\end{table}

Table~\ref{tab:incidentdurationfits} shows the results of the fitting procedure.
Strikingly, from the 1824 (segment, period)-pairs for which
an incident duration distribution needs to be estimated, 
there are only six pairs
for which none of the current fits is accepted.
Evidently, we could use more involved methods to find a 
proper fit for these six pairs.
Note that by the denseness of the class of phase-type 
distributions \citep[Section III.4]{Asmussen2003AppliedQueues},
we can include these into our Markovian framework as well.

\begin{remark}
{\em {In the above, we fit the distribution of incidents
based on their location and period of occurrence.
In case there is currently an incident in the network,
there is often additional information available, e.g.\ regarding
the nature of the incident, the involvement of trucks, etc.
Importantly, the presented fitting procedure is very flexible,
and can take this information into account. That is,
we may limit the data used to fit such that the considered
instances correspond with the current incident conditions.
For example, if it is known that there is currently a vehicle
breakdown, we may fit the distribution of the 
(residual) duration of this incident
on the subset of all incident
data entries occurred on the same segment,
having vehicle
breakdown as registered cause.}} \hfill $\Diamond$
\end{remark}

\begin{wrapfigure}{r}{.4\textwidth}
    \vspace{-10pt}
    \begin{minipage}{\linewidth}
    \centering\captionsetup[subfigure]{justification=centering}
    \includegraphics[width=\linewidth]{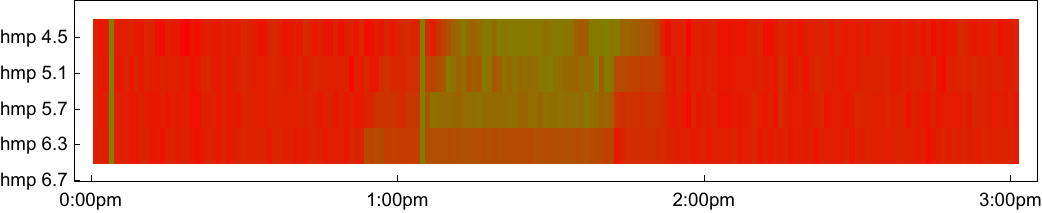}
    \subcaption{Heat map.}
    \label{fig:incidentspeedsHeat}
    \par\vfill
    \includegraphics[width=\linewidth]{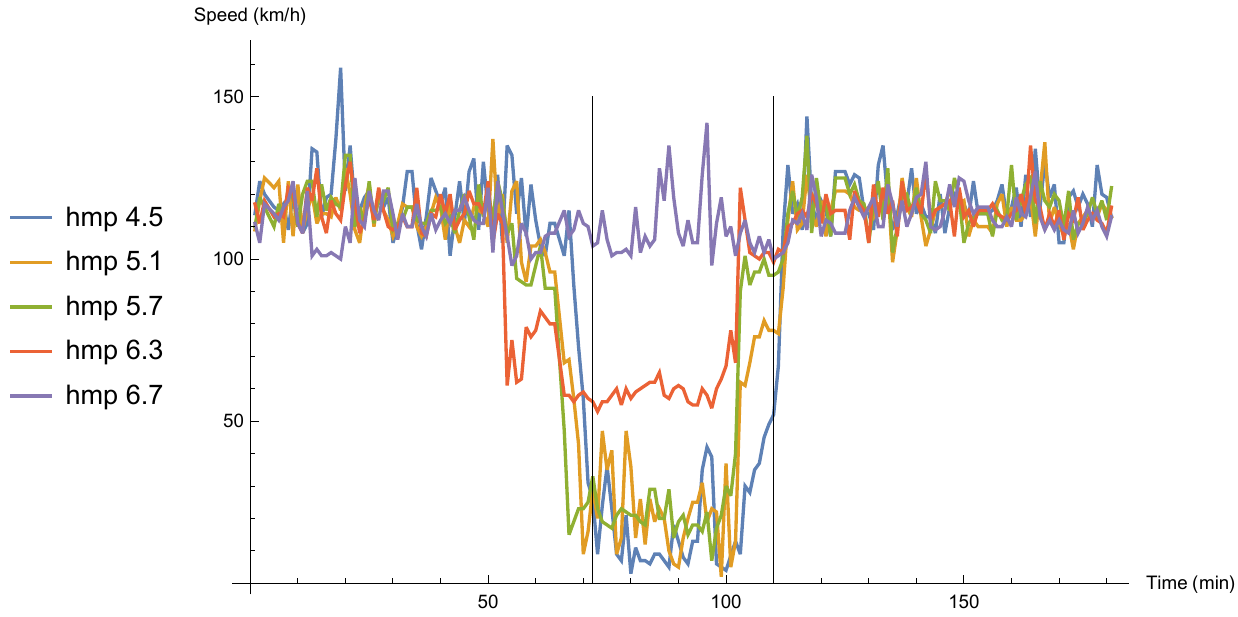}
    \subcaption{Line plot.}
    \label{fig:incidentspeedsLine}
    \end{minipage}
    \caption{Maximum speeds as registered by five
    detectors around an incident
    on \mbox{16-04-2017} at highway~A10. The left time
    point in both plots corresponds to midnight.
    The detectors are identified by the
    signalling hectometre indication
    closest to their location, with the
    indications being hectometer poles (hmp), 
    standing 100 metres apart through
    the highway network.}%
    \label{fig:incidentspeeds}%
\end{wrapfigure}
We proceed by investigating the speed patterns around
the reported incidents, such that we are able to model the impact
of incidents on the driveable vehicle speeds.
Importantly, we will argue that, during an incident,
the vehicle speeds on surrounding highway parts are generally 
well captured by one constant speed level, whose value depends
on the distance of the highway part to the location of the incident.
Naturally, this dependence is only local, and thus, for highway parts located far
from the incident, the driveable vehicle speed is not affected
by the incident.

Incidents are likely to lead to a substantial reduction of the driveable speed,
an example of which is given in Figure~\ref{fig:incidentspeeds}.
Note that, similar to the procedure for the non-incident speeds,
we show the maximum of the average driven speeds
over the road lanes, as this best reflects the driveable 
car speed. Now, we observe that 
for the longest part of the incident
depicted in Figure~\ref{fig:incidentspeedsLine}, the
speed levels at the individual detectors
are relatively stable. 
Indeed, with the two vertical lines indicating the reported start and end
time of the incident, we observe that, with exception
of the short periods during the start and end of the incident,
the vehicle speeds per detector fluctuate around a single speed level.
Thus, around every detector,
the driveable speed pattern during the studied incident 
could roughly be summarized by one speed value.
Using harmonic averaging, a representable incident speed
level for a slightly larger highway part, containing multiple
detectors, can be found. 

Figure~\ref{fig:incidentspeeds} also 
shows the spatio-temporal effect
of traffic jams: the incident
affects the velocities at detectors with
a further upstream distance from
the incident location typically somewhat later
and considerably less severe (in terms of the speed drop value).
Indeed, it can be observed that the speed level at
the detector around hectometer pole (hmp)~6.3 is only slightly affected by the incident, whereas 
the speed level at the detector around hmp~6.7
is the same before,
during, and after the incident.
Generally, an incident only affects
the speeds at highway parts relatively close to
the incident location. 
By studying historical
speed patterns of incidents, we can deduce the area
that is potentially affected by an incident located
at a given highway part.

In the above, we only showed the speed pattern during the incident
depicted in Figure~\ref{fig:incidentspeeds}. However,
the relatively low speed fluctuation during the largest part
of the incident does not only show
for this incident, but is a more observed phenomenon across the 
studied incidents in the Dutch highway network. 
Thus, we claim that, for every highway part
located around an incident, the
driveable speed is well described by just one speed level.
Moreover, conform the speed patterns in
Figure~\ref{fig:incidentspeeds},
typically, the highway parts that are not located
around an incident in the network, 
do not suffer a speed drop during the incident.

Now, recognizing the frequently
observed stability of speed patterns during incidents,
when working with the {\sc mvm} to model travel time
distributions, we can, for a given incident, simply use one 
speed level per highway link for all background
states encoding this incident.
Note that the observation
of low incident speed fluctuation is particularly useful 
in the case the incident is present
at the vehicle's departure. 
Indeed, in this case, the 
corresponding incident speed levels can directly be estimated by
the collected speeds in the minutes prior to the departure.

\section{Operationalizing the Markovian velocity model}\label{sec:oper}

Considering a vehicle that plans to traverse
a given path between an origin and a destination
in the Dutch highway network, at a specific day and time,
this section demonstrates how to employ the {\sc mvm} 
to obtain the corresponding travel time distribution.
The approach followed incorporates the effects of events 
that directly impact the driveable speeds. 
We consider the situation that
(i)~the analysis discussed in Section~\ref{sec:dataanalysis} has been performed, 
(ii)~the network state corresponding to the vehicle's departure is known
(in terms of the location and starting time of current incidents),
and (iii) information regarding 
existing or upcoming scheduled events 
(e.g. road work, bad weather conditions)
is available.

Recall that the {\sc mvm} models the events affecting
arc speeds through an environmental
background process $B(t)$.
Section~\ref{subsec:backgroundprocess} 
details how to construct this
background process $B(t)$, so as to
incorporate the recurrent and non-recurrent 
events potentially affecting
the departing vehicle's trip time,
and Section~\ref{subsec:speedlevels}
discusses how to set the driveable speed levels corresponding
to the different states of $B(t)$.

\subsection{Background process} \label{subsec:backgroundprocess}~\\
With $n = 1378$ directed links in the Dutch highway network,
the background process of the {\sc mvm} takes the form 
$B(t) = (Y(t),X_1(t),\ldots,X_n(t))$.
Here, $X_i(t)$ models the occurrence of incidents on link~$i$,
and~$Y(t)$ the \mbox{(semi-)predictable} events.
Concretely, $Y(t)$ captures
the recurrent, daily patterns with a Markov process~$Y_0(t)$, and
the scheduled events upon the vehicle's departure with Markov
processes $Y_1(t),\dots,Y_m(t)$ (in case there are~$m$ such events).
Importantly, with
incident characteristics e.g.\ dependent on the 
time of day, the Markov link processes~$X_i(t)$ 
are dependent on the state of the common process~$Y(t)$.

\subsubsection{Daily patterns}
Since the different traffic regimes during the day
are roughly described by the in Figure~\ref{fig:dependence}b identified periods,
$Y_0(t)$ captures the daily, recurrent traffic
patterns by modeling the durations of these six periods.
Observe that the durations of these periods 
are quite predictable.
Notably, including events whose durations have little
variability can be achieved by modeling 
these durations with Erlang phases.
Concretely, for given $k \in \mathbb{N}, t \in \mathbb{R}_{>0}$ and
$Z_1,\dots,Z_k$ i.i.d.\ exponentially distributed
random variables with mean $t/k$,
we have that $\sum_{i=1}^k Z_i$ is Erlang($k,k/t)$ distributed, such that
\begin{align*}
\mathbb{E}\Big[\sum_{i=1}^k Z_i \Big] = t, \quad \quad \quad
\text{Var}\Big[\sum_{i=1}^k Z_i \Big] = t^2/k.
\end{align*}
Thus, modeling predictable events with a given mean
by an Erlang-$k$ distribution with the same mean, we obtain a suitably
low variance when choosing $k$ sufficiently large.

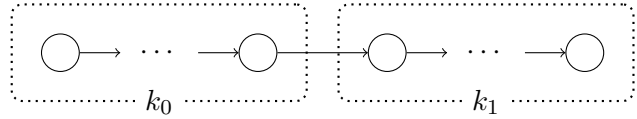
\begin{figure}[t]
\centering
\begin{tikzpicture}[node distance=1cm,auto]

\node[state, inner sep=5pt,minimum size=5pt, draw=none] (competitive) {};                           
\node (p11a) [state, inner sep=5pt,minimum size=5pt, below of=competitive, yshift=0cm] {};                           
\node (p12a) [state, inner sep=5pt,minimum size=5pt, right of=p11a, xshift=0.3cm, draw=none] {$\cdots$};           
\node (p13a) [state, inner sep=5pt,minimum size=5pt, right of=p12a, xshift=0.3cm] {};
\node (p21a) [state, inner sep=5pt,minimum size=5pt, right of=p13a, xshift=0.7cm] {};                           
\node (p22a) [state, inner sep=5pt,minimum size=5pt, right of=p21a, xshift=0.3cm, draw=none] {$\cdots$};
\node (p23a) [state, inner sep=5pt,minimum size=5pt, right of=p22a, xshift=0.3cm] {};

\draw[->] (p11a) -- (p12a);
\draw[->] (p12a) -- (p13a);
\draw[->] (p13a) -- (p21a);
\draw[->] (p21a) -- (p22a);
\draw[->] (p22a) -- (p23a);

\node[draw, thick, dotted, rounded corners, inner xsep=1em, inner ysep=1em, fit=(p13a) (p11a)] (box1a) {};
\node[fill=white] at (box1a.south) {$k_0$};
\node[draw, thick, dotted, rounded corners, inner xsep=1em, inner ysep=1em, fit=(p23a) (p21a)] (box2a) {};
\node[fill=white] at (box2a.south) {$k_1$};

\end{tikzpicture}
\caption{Example structure $Y_0(t)$.} 
\label{fig:structureZ2}
\end{figure}

Given the departure time of the vehicle, 
we know the remaining time of the current period, denoted by $t_0$,
as well as the lengths of the subsequent periods, 
denoted $t_1,t_2,\ldots,t_6$.
Now, in case duration $t_i$ is modeled with $k_i$ Erlang phases, 
the process $Y_0(t)$ in principle has
a total of as many as $k_0 + \dots + k_6$ states.
Fortunately,
travel times are typically in the order of minutes up to hours, so
that a trip will overlap with a low number of periods.
This concretely means that we only need to include into $Y_0(t)$ the phases
corresponding to these periods. 
Hence, with $M$ a crude upper bound for the time the
vehicle arrives at its destination, we may omit
all states belonging to periods that are
extremely unlikely to be entered before time $M$.
An example of a resulting structure of $Y_0(t)$
is given in Figure~\ref{fig:structureZ2},
only containing states belonging to
either of the first two periods.

\begin{figure}[t]
\centering
  \begin{subfigure}[b]{0.35\textwidth}
    \begin{tikzpicture}[node distance=1cm,auto]
\node (p11a) [state, inner sep=5pt,minimum size=5pt] {};                           
\node (p12a) [state, inner sep=5pt,minimum size=5pt, right of=p11a, xshift=0.3cm, draw=none] {$\cdots$};           
\node (p13a) [state, inner sep=5pt,minimum size=5pt, right of=p12a, xshift=0.3cm] {};
\node (p21a) [state, inner sep=5pt,minimum size=5pt, right of=p13a, xshift=1cm] {};

\draw[->] (p11a) -- node[above] {\tiny $\sfrac{k}{t'}$} (p12a);
\draw[->] (p12a) -- node[above] {\tiny $\sfrac{k}{t'}$} (p13a);
\draw[->] (p13a) -- node[above] {\tiny $\sfrac{k}{t'}$} (p21a);

\node[draw, thick, dotted, rounded corners, inner xsep=1em, inner ysep=1em, fit=(p13a) (p11a)] (box1a) {};
\node[fill=white] at (box1a.south) {$k$};
\node[draw, thick, dotted, rounded corners, inner xsep=1em, inner ysep=1em, fit=(p21a) (p21a)] (box2a) {};
\node[fill=white] at (box2a.south) {$1$};
\end{tikzpicture}
    \caption{Duration of an existing event modeled by
    an Erlang($k, k/t'$)-distribution.}
    \label{fig:ErlanginMVMa}
  \end{subfigure}
  \quad
  \begin{subfigure}[b]{0.55\textwidth}
\begin{tikzpicture}[node distance=1cm,auto]
\node (p11b) [state, inner sep=5pt,minimum size=5pt] {};                           
\node (p12b) [state, inner sep=5pt,minimum size=5pt, right of=p11b, xshift=0.3cm, draw=none] {$\cdots$};           
\node (p13b) [state, inner sep=5pt,minimum size=5pt, right of=p12b, xshift=0.3cm] {};
\node (p21b) [state, inner sep=5pt,minimum size=5pt, right of=p13b, xshift=1cm] {};
\node (p22b) [state, inner sep=5pt,minimum size=5pt, right of=p21b, xshift=0.3cm,draw=none] {$\cdots$};
\node (p23b) [state, inner sep=5pt,minimum size=5pt, right of=p22b, xshift=0.3cm] {};
\node (p31b) [state, inner sep=5pt,minimum size=5pt, right of=p23b, xshift=1cm] {};

\draw[->] (p11b) -- node[pos=0.6,above] {\tiny $\sfrac{k_1}{t'_1}$} (p12b);
\draw[->] (p12b) -- node[pos=0.3,above] {\tiny $\sfrac{k_1}{t'_1}$} (p13b);
\draw[->] (p13b) -- node[above] {\tiny $\sfrac{k_1}{t'_1}$} (p21b);
\draw[->] (p21b) -- node[pos=0.6,above] {\tiny $\sfrac{k_2}{t'_2}$} (p22b);
\draw[->] (p22b) -- node[pos=0.3,above] {\tiny $\sfrac{k_2}{t'_2}$} (p23b);
\draw[->] (p23b) -- node[above] {\tiny $\sfrac{k_2}{t'_2}$} (p31b);

\node[draw, thick, dotted, rounded corners, inner xsep=1em, inner ysep=1em, fit=(p13b) (p11b)] (box1b) {};
\node[fill=white] at (box1b.south) {$k_1$};
\node[draw, thick, dotted, rounded corners, inner xsep=1em, inner ysep=1em, fit=(p23b) (p21b)] (box2b) {};
\node[fill=white] at (box2b.south) {$k_2$};
\node[draw, thick, dotted, rounded corners, inner xsep=1em, inner ysep=1em, fit=(p31b) (p31b)] (box2b) {};
\node[fill=white] at (box2b.south) {$1$};
\end{tikzpicture}
    \caption{Time until and duration of a predicted event modeled by
    an Erlang($k_1, k_1/t'_1$) and Erlang($k_2, k_2/t'_2$)-distribution
    respectively.}
    \label{fig:ErlanginMVMb}
  \end{subfigure}
  \caption{Example structures scheduled events $Y_1(t),\dots,Y_m(t)$.}
  \label{fig:ErlanginMVM}
\end{figure}

We claim that using just a few phases 
per period (e.g.\ ,
$k_i \in \{5,\ldots,10\}$) is already sufficient to model
the period lengths well. 
The reason is that such a choice of $k_i$ already reduces the variance
of the time spent in period $i$ with a factor between 5 and 10 
compared to the exponential distribution. 
It is also noted that working with larger $k_i$ values would
ignore the intrinsic fluctuations of the periods' start and end times. 
Moreover, working
with large values of $k_i$ has the
undesired consequence of inflating
the  state space of $B(t)$, thus leading to
a high computational complexity.

\subsubsection{Scheduled events}
We can use the same ideas to capture the duration of 
the $m$ scheduled events 
that are modeled through the processes $Y_1(t),\dots,Y_m(t)$.
That is, if event~$i$ is an existing event (i.e.,
present at the vehicle's departure) for which
the expected remaining duration is known to equal $t'$,
we can use the Erlang($k, k/t'$)-distribution to
model the duration of event~$i$ 
(Figure~\ref{fig:ErlanginMVMa}). 
In case event~$i$ is 
not an existing but a forecasted event, $Y_i(t)$ should,
besides the duration of the event, also
include Erlang phases that model the time until the start
of the event
(Figure~\ref{fig:ErlanginMVMb}).

\begin{remark}
{\em {If, besides information on the mean duration 
of a scheduled event (or the time until its start), there is information available 
on the {\it variance} of the duration (or the time until its start), one can alternatively fit its distribution
with the two-moment phase-type matching techniques that were presented in Section~\ref{subsec:incidents}.
}} \hfill $\Diamond$
\end{remark}

\subsubsection{Incidents}
With the general structure of $Y(t)$ known, we are
now able to characterize the Markov processes
$X_1(t),\dots,X_n(t)$, that model the incidents
on the links of the network, conditional on the state of $Y(t)$.
Concretely, we let the dynamics of
$X_i(t)$ depend on the state of $Y_0(t)$, since
it was concluded in Section~\ref{sec:dataanalysis} that
incident dynamics depend on the
specific period of the day.
Note that these incident dynamics should cover the incident
duration itself, as well as the
inter-incident time. Recall that in Section~\ref{sec:dataanalysis},
these distributions are fitted per highway \textit{segment} (i.e.,
highway part between two highway intersections), whereas
$X_i(t)$ should capture these distributions per highway \textit{link}
(i.e., highway part between two ramps).

Given the period of the day, encoded by the state the process $Y_0(t)$ is in, 
we have shown in Section~\ref{subsec:interincident} that, for every highway segment, 
the time between two incidents in this period can be modeled 
by an exponential distribution.
We will now argue that the inter-incident duration
on the links that partition this segment can be described
by the exponential distribution as well.
This implies that, for a link $i$, the process $X_i(t)$
contains just one exponential state that represents the situation
in which the link is incident-free.
The mean time spent in this state depends on the period
of the day, i.e., on the state of $Y_0(t)$.

Denote by $1/\lambda_{j,k}$ the mean inter-incident time on segment $j$
in case $Y_0(t) = k$ (i.e., $\lambda_{j,k}$ is the rate of the corresponding exponential distribution). 
To see that the inter-incident distribution
of the links that partition this segment is indeed
exponential, note that the initiation of an incident on segment~$j$ corresponds
to the initiation of an incident on \textit{one} of these links.
Therefore, we assign a value~$p_i^j \in [0,1]$ to every link~$i$
on segment~$j$, representing the probability
that, given there is an incident on segment~$j$, this incident
has occurred at link $i$.
As a natural proxy for $p_i^j$ we take the ratio of the lengths of link~$i$ and segment~$j$.
Observe that, in modeling terms, the inter-incident
time on link~$i$ will have an 
Exponential($p_i^j\lambda_{j,k}$)-distribution.
Importantly, since the inter-incident time on segment~$j$ is the minimum
of the inter-incident times on the links at $j$, 
we (consistently) obtain the Exponential($\lambda_{j,k}$)-distribution
for the inter-incident time on the full segment.

If, for link $i$, the process $X_i(t)$ transitions
out of the inter-incident state, this corresponds
to the occurrence of an incident on this link. 
As was concluded in Section~\ref{subsec:incidents},
the distribution of the duration 
of such an incident depends on the period of
the day in which the incident occurs.
Therefore, for every period modeled
by $Y_0(t)$, the process $X_i(t)$ should contain states
that describe the duration of an incident
which started in that period.
For example, in case $Y_0(t)$ is structured as 
in Figure~\ref{fig:structureZ2}, $X_i(t)$ contains
states that describe an incident which started in the
period modeled by $k_0$ phases, as well as states
that describe an incident which started in the
period modeled by $k_1$ phases.

Given the period that corresponds
to the state of $Y_0(t)$, we have fitted
the distribution of the duration of an incident 
starting in this period
for every highway segment
in the Dutch network (Section~\ref{subsec:incidents}).
Now, given such a segment, we will use the
corresponding incident distribution
for every link that partitions this segment.
Thus, if a highway segment between two
intersections consists of three links, separated by
ramps, the incident distribution that was fitted
for the segment is used to model incidents
on all these three links.
Recall that, because the fitted distributions all 
fall in the category of phase-type distributions,
we can directly include these
in the Markov processes $X_i(t)$.

A special case are the incidents that
are already present upon the vehicle's departure.
Given that link $i$ has an incident, knowledge of the starting time of
the incident yields both the distribution of the incident length
and the current running time of the incident, from
which we can deduct the distribution of the remaining
incident length. Then, the process $X_i(t)$ should also
include states modeling this remaining incident length,
which $X_i(t)$ visits before transitioning to the inter-incident state.
Trivially, if the incident distribution is exponential, 
the remaining incident time is also exponentially distributed.
For an incident with current running time $t > 0$ and
hyperexponentially distributed length with parameters $p \in [0,1], \mu_1, \mu_2
\in \mathbb{R}_{>0}$, we have:
\begin{align*}
    \mathbb{P}(X\!>\!t\!+\!s\,|\,X\!>\!t) 
    = \frac{\mathbb{P}(X\!>\!t\!+\!s)}{\mathbb{P}(X\!>\!t)}
    = \frac{pe^{-\mu_1(t+s)} + (1\!-\!p)e^{-\mu_2(t+s)}}{pe^{-\mu_1t} + (1\!-\!p)e^{-\mu_2t}} = qe^{-\mu_1 s} + (1\!-\!q)e^{-\mu_2 s},
\end{align*}
with \[q := \frac{pe^{-\mu_1 t}}{pe^{-\mu_1t} + (1\!-\!p)e^{-\mu_2t}}.\]
Thus, the remaining incident time has a hyperexponential distribution as
well, with parameters $q, \mu_1, \mu_2$.
It can be proven that the distribution of an Erlang$(k,\mu)$ random variable, conditioned on being at least $t$, is
a mixture of Erlang$(j,\mu)$ distributions, with
$j=1,\dots,k$ (Theorem~\ref{thm:conditionalErlang}, Appendix~\ref{appendix:conditionalErlang}). 
This gives that the
remaining time of an Erlang-2 distribution
can be cast in the Markovian framework.
Moreover, it can now easily be deduced that
the remaining incident time of a mixture
Erlang distribution is a mixture Erlang distribution
as well.

\subsection{Speed levels} \label{subsec:speedlevels}~\\
To obtain the travel time distribution for the vehicle,
the speed levels corresponding to the different background
states have to be specified. That is, for all~$i$
and all $s \in S$, we need to set a
value for $v_{a_i}(s)$, the
speed at which vehicles are moving on link $a_i$
given $B(t) = s$. Without loss of generality,
we focus on the speed levels of link $a_1$.
Recall from Section~\ref{subsec:incidents} that,
with incidents being local events, only incidents
on links surrounding $a_1$ will affect the speed
on this link. Denote this set of links whose congestion status
affects $a_1$ by $A_{a_1}$.

Let $s \in S$ be a state that corresponds to 
a setting in which the arcs in $A_{a_1}$ 
are incident-free, and there are no existing
scheduled events in the network. Then,
the driveable speed on $a_1$ is fully determined
by the daily velocity patterns. 
The speed and flow data analysis as 
performed in Section~\ref{subsec:interincident} has revealed
during which periods, and thus for which states of $Y_0(t)$, 
the driveable speed level on $a_1$ 
equals the free-flow speed.
In case the state of $Y_0(t)$ belongs
to a non free-flow (but still relatively constant) speed period, 
historical averaging
of the maximum of the road lane speeds yields
a representative speed level per loop detector
on link $a_1$.
For this averaging, 
we only use speed data of the days of the week
that match the current day, and only from 
a few weeks preceding the vehicle's departure.
This way, we account not only for within-day time-dependence,
but for dependence on the day of the week and season as well.
Now, to obtain a representative speed level for the complete link,
we simply take the weighted harmonic mean of the speeds levels
of the individual detectors located on the link.
The weights are set to account for the non-uniform placement of the loops
on the link, the weight
of a single detector being the total distance
between the midway points to its neighbors,
or, in case of only one neighbor, the distance
from this one midway point to the boundary of the link.

Importantly, 
we can easily work with more elaborate 
speed patterns, in case one constant speed
level is not representative for the 
speed pattern in the considered period. 
That is, by assigning different speed levels
to the different Erlang phases that model the duration
of this period, the driveable speed (as function
of time) is a step function. Being able to model stepwise
speed functions allows to replicate more complex daily patterns,
such as the observed V-shape during the rush hour on 
the A10 (Figure~\ref{fig:AvMaxSpeedsA10}).
The numerical examples in
Section~\ref{sec:numeric} will indeed reveal
that it is not always necessary to replicate
these shapes perfectly, and that,
for the considered instances, working
with step functions 
is already satisfactory.

Now, let $s \in S$ be a state that corresponds to a setting
in which one of the arcs in~$A_{a_1}$ is not incident-free.
As argued in Section~\ref{subsec:incidents}, 
during this incident, the driveable speed
on link~$a_1$ can roughly be 
described by one speed level.
In case the incident is already
present upon the vehicle's departure, real-time information
regarding the driven speeds at the detectors on~$a_1$
may be available. 
If these reveal that the incident is in a
stationary state, i.e., speed
fluctuations in the minutes prior to the
departure are only mild, we can
set the speed level on~$a_1$ in state~$s$ as the
current speed level on~$a_1$.
Alternatively, if the last minutes of speed 
data do not show a somewhat 
stable pattern, the link speed that corresponds
to the last minute of data is 
set as speed level for $s \in S$.
In case the incident has just started or is in the
process of clearance, typically corresponding with respectively
a decreasing or increasing speed trend,
this estimate is expected to be a
better representative than an 
average over a longer period.

\begin{figure}[ht]
    \centering
    \begin{subfigure}[b]{0.35\textwidth}
    \includegraphics[width=\textwidth]{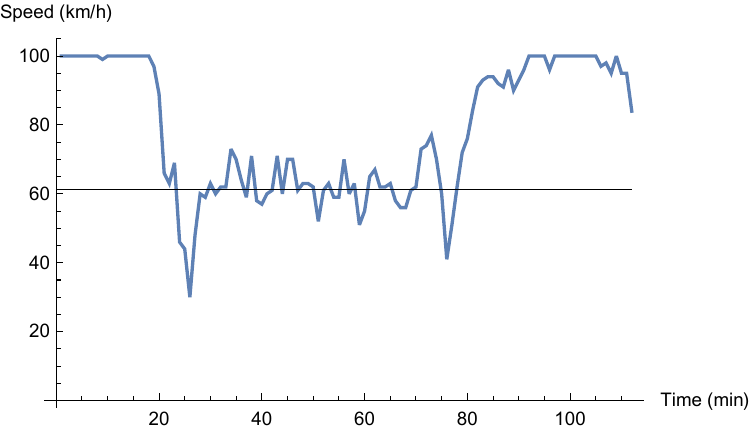}
    \end{subfigure}
    \quad \quad
    \begin{subfigure}[b]{0.35\textwidth}
    \includegraphics[width=\textwidth]{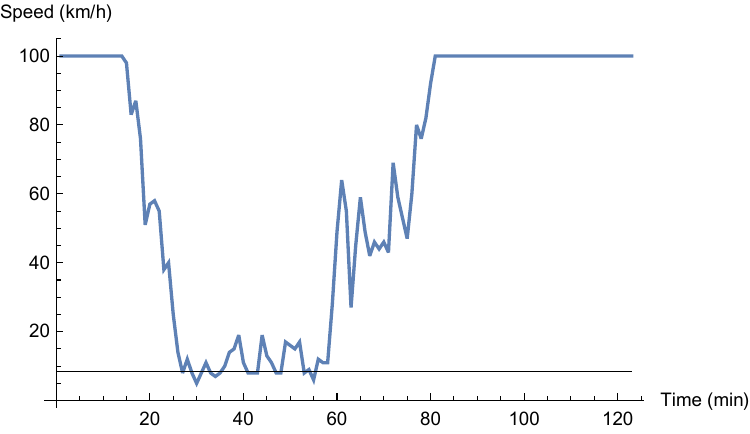}
    \end{subfigure}
    \caption{Identifying the stationary
    speed level of two historical incidents.}
    \label{fig:stablespeedpattern}
\end{figure}

In case there is no real-time speed
data available, or the incident is not already present
at the vehicle's departure, 
the speed level of the incident is unknown,
and estimated by the average of
the historical speed levels of incidents
located on the same link.
To obtain the stable speed level of a historical
incident, we propose to take, as above, the weighted harmonic
average of the stable speed patterns of the individual 
loop detectors.
For every such detector, the stable speed level is
identified by computing, for every ten minutes of
data around the time interval of the incident,
the mean and variance in registered speeds.
Then, from all means below 80 km/h,
the stable speed level is set as
the one with the minimum variance.
Figure~\ref{fig:stablespeedpattern} shows the result for two historical
incidents on the highway A10.

Evidently, in case there are multiple incidents
in the vicinity of arc~$a_1$,
the speed on~$a_1$ is at least as much
affected as it would be by one of these individual incidents.
Therefore, for $s \in S$ for which there are
multiple arcs in~$A_{a_1}$ that have incidents, 
we simply set the velocity on~$a_1$ 
in state~$s$ as the minimum
of the speed levels of~$a_1$ 
corresponding to the individual incidents.

In case there are existing or upcoming scheduled
events modeled through $Y(t)$ (e.g.\ road work or 
bad weather), the effect of these
events should be taken into account as well.
The speed levels on the arcs affected by the scheduled
events are estimated in a similar way as the incident
speeds. That is, if the event is present at the vehicle's
departure and current speeds are available, these speeds
are used to estimate the presentable speed level.
If these speeds are not available, or the event
is a future event, historical speed data is used
to determine the correct speed level. 

\begin{figure}[ht]
    \centering
    \includegraphics[width=0.35\textwidth]{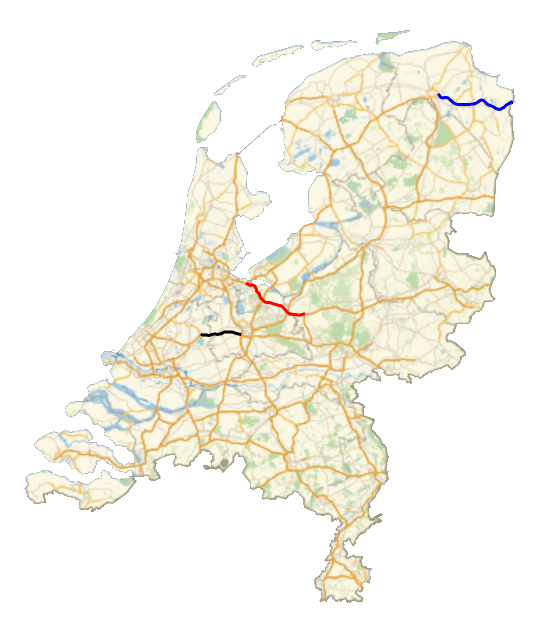}
    \caption{Three considered paths in the Dutch highway network.}
    \label{fig:threeroutes}
\end{figure}

\section{Numerical experiments} \label{sec:numeric}

Having described and substantiated
the individual components of our procedure,
we will now display the resulting
travel time distributions
for several case studies
and provide 
examples of the advantage of our
model as compared to traditional
travel time estimation methods.
The case studies consider the
traversal of three west-to-east directed paths in
the Dutch highway network,
depicted in Figure~\ref{fig:threeroutes},
under various traffic scenarios. Concretely,
for each of these paths, we look at the travel time distribution 
of vehicles traversing the path in the non-rush
hour and rush hour setting, in case the path is incident-free
upon departure (Section~\ref{subsec:ttdistr}). Additionally, in Section~\ref{subsec:ttdistrinc},
we consider a traffic setting where, 
upon the vehicle's departure,
there is an incident on the path to travel.
With the time until clearance of the incident
a random variable, our procedures, taking
this uncertainty into account, are shown to
outperform deterministic estimations.

Before doing so, we briefly explain
why the available travel time data was not adequate
for a more detailed (numerical) assessment of the performance 
of our methodology.
That is, naturally, one would want to compare the travel 
time distributions we obtain for the traversal of the three paths under
various traffic settings with travel time data from
the Dutch highway network. However, hampered by the availability and 
quality of the travel time data as provided by
{\sc ndw}, such a comparison could
not be performed.

Limitations arise due to the fact that, with poor availability of both floating car data and travel time data collected via Bluetooth or cameras (for the years of study), the {\sc ndw} data only contains rough,
rounded estimates of average travel times, based on measurements with loop detectors. In fact, per trajectory and per minute, there is only one {\sc ndw} 
data value that represents the \emph{general mean travel time}, averaged over all vehicles (cars and trucks) on the segment under consideration. Since the maximum speed trucks are allowed to drive in The Netherlands is lower than the maximum 
car speed, a comparison would (incorrectly) lead to the conclusion that our travel-time estimates are too low. 
Indeed, with traffic heterogeneity playing a 
prominent role, the realized speed levels are typically below
the actual driveable speeds, limiting a fair numerical comparative analysis. 
A second conceptual complication is that our procedures are based on capturing the travel times that vehicles are \emph{effectively} able to drive, in contrast to the collected travel time data, which only reflects (rough estimates of) \emph{realized} travel times, which are subject to the heterogeneity in driving style of individual vehicles. 

\subsection{Travel time distributions under the absence of incidents}\label{subsec:ttdistr}~\\
Focusing on the travel time distributions
on the three paths of Figure~\ref{fig:threeroutes},
it is important to note that the paths are of a different
characteristic nature. For example, they
differ greatly in terms of incident-proneness,
as can be observed from Figure~\ref{fig:interincloc}. 
For the blue path, which is approximately
37 km long and located in a rural area, the mean inter-incident time is highest.
The red path, which is approximately 30 km long and brings
vehicles from the city of Amsterdam to a more rural area,
has a relatively low mean inter-incident time.
The mean time between incidents is lowest
for the circa 38 km long black path between
two busy urban areas.

\begin{figure}[b]
\centering
\begin{subfigure}{.3\textwidth}
  \centering
  \includegraphics[width=.9\linewidth]{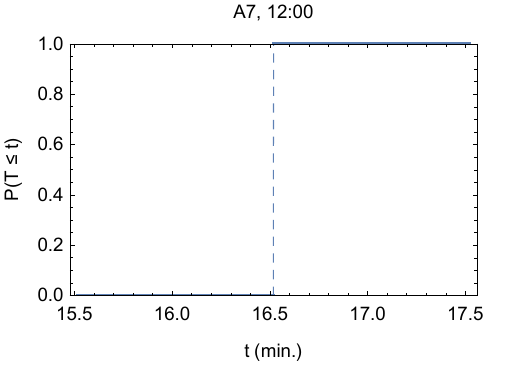}
  \caption{A7}
  \label{fig:nonrushA7}
\end{subfigure}%
\begin{subfigure}{.3\textwidth}
  \centering
  \includegraphics[width=.9\linewidth]{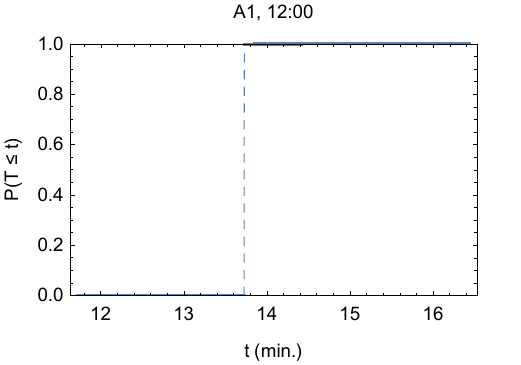}
  \caption{A1}
  \label{fig:nonrushA1}
\end{subfigure}
\begin{subfigure}{.3\textwidth}
  \centering
  \includegraphics[width=.9\linewidth]{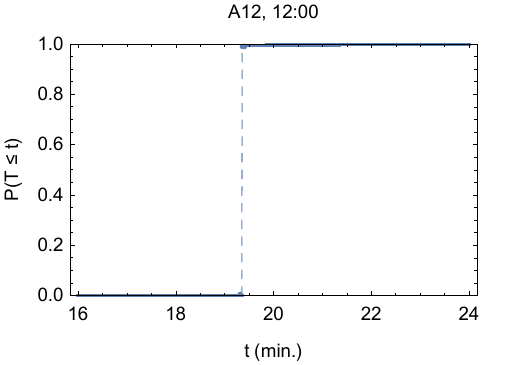}
  \caption{A12}
  \label{fig:nonrushA12}
\end{subfigure}
\caption{Cumulative travel time distribution estimates for departure at a regular 
Wednesday at noon in 2019.}
\label{fig:nonrush}
\end{figure}

We first consider the traversal of the 
paths in an incident-free non-rush hour
setting. Specifically, we look at vehicles
traveling the paths on a regular 
Wednesday (i.e., no 
school holiday or national holiday) in the year 2019
at noon, in case there are
no registered incidents upon departure.
In accordance with Section~\ref{sec:oper},
the speed levels in the non-incident setting
are estimated by averaging over the speeds
of the four regular and incident-free Wednesdays prior
to the considered day, and
the speed levels of future incidents are estimated by averaging  
speed levels of historical incidents on the same highway link.
The resulting cumulative travel-time distributions are
presented in Figure~\ref{fig:nonrush}.
Recall that our model does not take into account fluctuations that typically arise due to the differences in driving styles, which explains the nearly deterministic pattern.
Indeed, since both the probability of incident occurrence during
the trip and the probability of hitting the next time period,
corresponding to rush hour conditions,
are extremely small, the driveable vehicle speed during
the trip is well described by one constant velocity level.

\begin{figure}[t]
\centering
\begin{subfigure}{.3\textwidth}
  \centering
  \includegraphics[width=.9\linewidth]{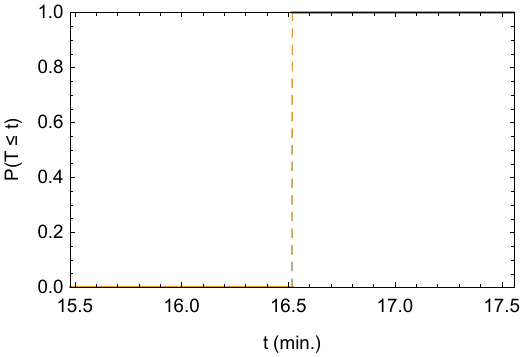}
  \caption{A7}
  \label{fig:rushA7}
\end{subfigure}%
\begin{subfigure}{.3\textwidth}
  \centering
  \includegraphics[width=.9\linewidth]{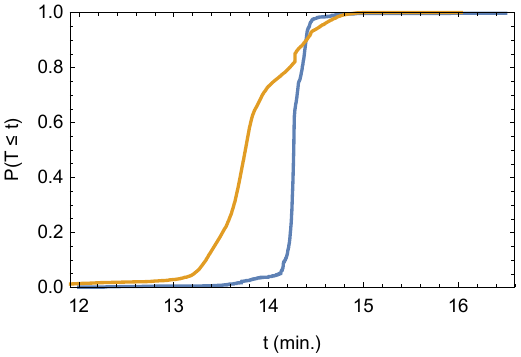}
  \caption{A1}
  \label{fig:rushA1}
\end{subfigure}
\begin{subfigure}{.3\textwidth}
  \centering
  \includegraphics[width=.9\linewidth]{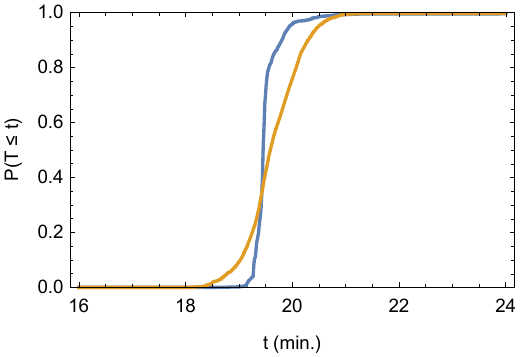}
  \caption{A12}
  \label{fig:rushA12}
\end{subfigure}
\caption{Cumulative travel time distribution estimates for departure at Tuesday 3:15 p.m.\ (blue)
and 6:00 p.m.\ (yellow) in 2019.}
\label{fig:rush}
\end{figure}

Now, let us alternatively consider a vehicle
that departs at a regular 2019 Tuesday at 3:15~p.m.\ or 6:00~p.m.
At the first time instant, upon the vehicle's departure,
the onset of rush hour is in the near future, whereas
the second departure instant falls 
within the rush hour period. 
In contrast to the non-rush hour setting,
the travel time distributions at these instances,
as displayed in Figure~\ref{fig:rush}, clearly show
the different characteristic natures of the considered
paths. That is, with low daily flow levels in 
all periods, the driveable
speed levels on the A7 highway equal the free flow speed,
again leading to an approximately deterministic distribution,
independent of the departure time.
On the other hand, the A1 and A12 highway do show uncertainty.
For departure at 3:15 p.m., this uncertainty is
only mild, as a large part of the paths is traversed in
non-rush hour setting, which knows constant traffic speeds.
However, the width of the travel time distributions
is larger for departure at 6:00 p.m., with the travel
times suffering from the (semi-)random onset of the
different rush-hour speed trends.

\subsection{Travel time distributions under the presence of incidents}\label{subsec:ttdistrinc}
\begin{figure}[ht]
    \centering
    \includegraphics[width=0.4\textwidth]{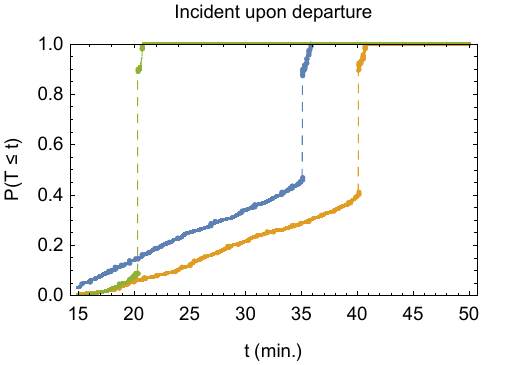}
    \caption{
    Travel time distribution when traversing the A1 path
    in incident setting. Departure time is after
    $25\%$ (blue), $50\%$ (yellow) or
    $75\%$ (green) of the total incident duration.}
    \label{fig:incidentA1}
\end{figure}

~\\
Due to the relatively high amount of
uncertainty, the most interesting distribution estimates arise in case,
upon the vehicle's departure, there is an incident
on the path to travel. 
In such traffic scenarios,
our procedures are clearly beneficial
over traditional, deterministic methods that either restrict
to working with current speeds or only take recurrent patterns
into account. That is, these are unable
to work with random future changes in traffic
conditions, yielding poor travel
time estimations in case there is a high
probability of such changes.
For example, traditional methods
are often insufficient when the incident is located
at the end of the path to travel, since, with relatively
much time until the vehicle reaches the incident location,
there is typically a high probability of incident
clearance before reaching the incident. 
Another example in which 
current route-planners may perform unsatisfactorily
is presented in Figure~\ref{fig:incidentA1},
which shows the cumulative travel-time distributions for 
a specific incident, a defective truck at the first
part of the A1 path, when departing after 
25\%, 50\% or 75\% of the total
reported incident duration. 

To illustrate how the {\sc MVM} can be used to improve
routing advice, we first consider a vehicle 
departing at the 25\% time instant.
Obviously, the remaining
incident duration is unknown to the driver itself. 
With traditional methods not accounting for random changes, 
route planners will predict a travel time of 
35.1 minutes, i.e., the travel time that corresponds to the location of the probability mass point of the 25\% distribution.
Indeed, observe that the mass point corresponds to the scenario in which the
considered incident is not cleared during the traversal of the links
affected by this incident.
In contrast, our method \textit{does} take into account that with a certain probability, the incident will have cleared before the vehicle arrives 
at the congested links. 
On the A1 path,
there is a high percentage of reported incidents
with relatively short duration.
Thus, the distribution of the remaining incident
duration has high probability mass on the left side, 
yielding a high probability of clearance
before reaching the incident. The mass
point of the obtained distribution indeed
reveals that there is an almost 40\% chance
that the incident has cleared before the vehicle arrives.
Therefore, in expectation, the travel time will
be significantly less than the 35.1 minutes
estimated by traditional route-planners.

In the case where 50\% percent of the incident 
duration has elapsed upon departure, it can be observed
from Figure~\ref{fig:incidentA1} 
that our procedure estimates that there is a high 
probability that the incident is cleared during
the traversal of the path, yielding,
again, an expected travel time that is significantly less than 
the 40.1 minutes that will be estimated by 
traditional methods. Note that, as the position of the mass point
provides an impression for the incident speed levels,
the driveable incident speeds are lower than those recorded
after 25\% of the incident. This can be explained by
the fact that, after 25\% of the total incident duration, 
the speed levels at detectors further away from the incident may not be
affected yet, since the traffic jam is
still accumulating, leading to slightly lower 
travel time values when compared to
the 50\% scenario.

When departing after 75\% of the incident duration,
the mean travel time estimate of the {\sc mvm} is almost the
same as when using a deterministic method: they would both
yield an estimate of 
20.3 minutes.
This is inherent to the small probability
that the incident is 
cleared during traversal
of the path. Observe, moreover, that the
location of the mass point indicates that,
compared to departure after 25\% and 50\% of the incident duration, the 
speeds are significantly closer to their non-incident version. 
Reviewing the incident characteristics, it is
revealed that this is due to the fact that 
the lanes that were closed at the 25\% and 50\% 
instances, are fully opened after 75\%, with recovery 
speeds shortly revealing a shockwave pattern.

\section{Conclusive remarks}

In this paper we presented comprehensive techniques to describe the randomness
of incidents in a highway network, in terms of their frequency, duration and impact
on vehicle speeds. 
With these results, we were able to operationalize the Markovian velocity
model, a stochastic model that tracks both recurrent
and non-recurrent traffic events that affect driveable vehicle speeds.
Numerical experiments demonstrated the impact of recurrent and non-recurrent effects
on such travel time distribution estimates in various traffic settings.

We have shown that, on a given highway segment,
both the incident duration and inter-incident time are dependent 
on the time of day, but that we can deal with this time-dependence
by working with periods in which these effects are essentially constant.
For every highway segment, the inter-incident time within each
of these periods is well described by an exponential distribution, 
whereas, in nearly all cases, the duration of an incident starting in this period 
fits a phase-type distribution with a relatively low number of phases.
When fitting the incident data, we have used the collection of all registered
incidents per highway segment, and not distinguished on environmental conditions. 
A future study could include both incident
and weather data, and investigate the impact of different weather
conditions on the incident length and driveable vehicle speeds.
This could further improve the prediction results in case,
upon the vehicle's departure, weather conditions are poor.

To operationalize the Markovian velocity model, 
we presented methods to obtain representative levels
for the driveable vehicle speeds in both the incident and inter-incident setting.
In the inter-incident setting, it could be observed that these speed
levels depend on the period-of-day, day-of-week and time-of-year.
To tackle these dependencies, we proposed a simple, fast and transparent
clustering method, in which we just average over the speeds
observed in the same period and on the same day, in the weeks
previous to the vehicle's departure. Evidently, more enhanced 
prediction methods could be used to find representative speed levels.

The numerical experiments we conducted showed the impact of 
recurrent traffic patterns,
current incidents and potential future incidents on the travel
time distribution estimates. It was observed that the impact
of future incidents is minor, whereas the
impact of both rush hour and current incidents
is more pronounced. Future work could be specified
towards incorporating the impact of second-order effects
into the travel time distribution estimates as well.
A potential suggestion would be to incorporate the
heterogeneity in driving by style by letting the vehicle
speeds -- instead of being constant -- be described by a distribution
that depends on the state of the background process.

As discussed in Section~5, the absence of reliable 
travel time data prohibits a full comparison between the obtained
travel times estimates and real-world data. 
However, we have been able to show the advantages compared to traditional travel-time prediction methods through some illustrative examples.
A more extensive comparison is clearly desirable, and should be carried out once
there is access to more suitable travel time data. Note that, 
with current technical advances, floating car data is expected
to become available on a large scale in the upcoming years.

\bibliographystyle{agsm}
\bibliography{references.bib}

@misc{ndwdatabase,
    title = {{National Road Traffic Data Portal}},
    year = {2020},
    author = {{NDW}},
    url = {https://www.ndw.nu/}
}

@misc{rwsdatabase,
    title = {{Traffic Jam Data}},
    year = {2015},
    author = {{Rijkswaterstaat}},
    url = {https://www.rijkswaterstaat.nl/apps/geoservices/geodata/dmc/filedata/}
}

@article{Levering2022AConditions,
    title = {{A framework for efficient dynamic routing under stochastically varying conditions}},
    year = {2022},
    journal = {Transportation Research Part B: Methodological},
    author = {Levering, Nikki and Boon, Marko and Mandjes, Michel and N{\'{u}}{\~{n}}ez-Queija, Rudesindo},
    month = {6},
    pages = {97--124},
    volume = {160},
    doi = {10.1016/j.trb.2022.04.001},
    issn = {01912615}
}

@article{Javid2018AData,
    title = {{A framework for travel time variability analysis using urban traffic incident data}},
    year = {2018},
    journal = {IATSS Research},
    author = {Javid, Roxana J. and Javid, Ramina J.},
    number = {1},
    month = {4},
    pages = {30--38},
    volume = {42},
    publisher = {Elsevier B.V.},
    doi = {10.1016/j.iatssr.2017.06.003},
    issn = {03861112}
}

@article{Khattak1995ADuration,
    title = {{A Simple Time Sequential Procedure For Predicting Freeway Incident Duration}},
    year = {1995},
    journal = {IVHS Journal},
    author = {Khattak, Asad J and Schofer, Joseph L and Wang, Mu-Han},
    number = {2},
    pages = {113--138},
    volume = {2}
}

@article{Chung2012AnalyticalAnalysis,
    title = {{Analytical method to estimate accident duration using archived speed profile and its statistical analysis}},
    year = {2012},
    journal = {KSCE Journal of Civil Engineering},
    author = {Chung, Younshik and Yoon, Byoung-Jo},
    number = {6},
    month = {9},
    pages = {1064--1070},
    volume = {16},
    doi = {10.1007/s12205-012-1632-3},
    issn = {12267988}
}

@article{Chen2020AnalyzingData,
    title = {{Analyzing travel time distribution based on different travel time reliability patterns using probe vehicle data}},
    year = {2020},
    journal = {International Journal of Transportation Science and Technology},
    author = {Chen, Zhen and Fan, Wei},
    number = {1},
    month = {3},
    pages = {64--75},
    volume = {9},
    publisher = {Elsevier B.V.},
    doi = {10.1016/j.ijtst.2019.10.001},
    issn = {20460449}
}

@book{Asmussen2003AppliedQueues,
    title = {{Applied Probability and Queues}},
    year = {2003},
    author = {Asmussen, Soeren},
    edition = {Second},
    publisher = {Springer New York}
}

@article{Chen2019DataPrediction,
    title = {{Data analytics approach for travel time reliability pattern analysis and prediction}},
    year = {2019},
    journal = {Journal of Modern Transportation},
    author = {Chen, Zhen and Fan, Wei},
    number = {4},
    month = {12},
    pages = {250--265},
    volume = {27},
    publisher = {Springer},
    doi = {10.1007/s40534-019-00195-6},
    issn = {21960577}
}

@inproceedings{Xie2020DeepPrediction,
    title = {{Deep Graph Convolutional Networks for Incident-Driven Traffic Speed Prediction}},
    year = {2020},
    booktitle = {CIKM '20: Proceedings of the 29th ACM International Conference on Information {\&} Knowledge Management},
    author = {Xie, Qinge and Guo, Tiancheng and Chen, Yang and Xiao, Yu and Wang, Xin and Zhao, Ben Y.},
    pages = {1665--1674},
    publisher = {},
    isbn = {9781450368599},
    doi = {10.1145/3340531.3411873}
}

@article{Kharoufeh2004DerivingProcesses,
    title = {{Deriving Link Travel-Time Distributions via Stochastic Speed Processes}},
    year = {2004},
    journal = {Transportation Science},
    author = {Kharoufeh, Jeffrey P and Gautam, Natarajan},
    number = {1},
    pages = {97--106},
    volume = {38}
}

@article{Guner2012DynamicInformation,
    title = {{Dynamic routing under recurrent and non-recurrent congestion using real-time ITS information}},
    year = {2012},
    journal = {Computers and Operations Research},
    author = {G{\"{u}}ner, Ali R. and Murat, Alper and Chinnam, Ratna Babu},
    number = {2},
    month = {2},
    pages = {358--373},
    volume = {39},
    doi = {10.1016/j.cor.2011.04.012},
    issn = {03050548}
}

@article{Garib1997EstimatingDelays,
    title = {{Estimating Magnitude and Duration of Incident Delays}},
    year = {1997},
    journal = {Journal of Transportation Engineering},
    author = {Garib, Atef and Radwan, A Essam and Al-Deek, Haitham},
    number = {6},
    pages = {459--466},
    volume = {123}
}

@article{Guessous2014EstimatingConditions,
    title = {{Estimating travel time distribution under different traffic conditions}},
    year = {2014},
    journal = {Transportation Research Procedia},
    author = {Guessous, Younes and Aron, Maurice and Bhouri, Neila and Cohen, Simon},
    pages = {339--348},
    volume = {3},
    publisher = {},
    doi = {10.1016/j.trpro.2014.10.014},
    issn = {23521465}
}

@article{Filipovska2021EstimationCorrelations,
    title = {{Estimation of Path Travel Time Distributions in Stochastic Time-Varying Networks with Correlations}},
    year = {2021},
    journal = {Transportation Research Record},
    author = {Filipovska, Monika and Mahmassani, Hani S and Mittal, Archak},
    number = {11},
    pages = {498--508},
    volume = {2675},
    doi = {10.1177/03611981211018464}
}

@article{Bergel-Hayat2013ExplainingEffects,
    title = {{Explaining the road accident risk: Weather effects}},
    year = {2013},
    journal = {Accident Analysis {\&} Prevention},
    author = {Bergel-Hayat, Ruth and Debbarh, Mohammed and Antoniou, Constantinos and Yannis, George},
    pages = {456--465},
    volume = {60},
    doi = {https://doi.org/10.1016/j.aap.2013.03.006},
    issn = {0001-4575}
}

@article{Giuliano1989IncidentFreeway,
    title = {{Incident characteristics, frequency, and duration on a high volume urban freeway}},
    year = {1989},
    journal = {Transportation Research Part A: General},
    author = {Giuliano, Genevieve},
    number = {5},
    pages = {387--396},
    volume = {23},
    doi = {https://doi.org/10.1016/0191-2607(89)90086-1},
    issn = {0191-2607}
}

@article{Qiu2021MachineAnalyses,
    title = {{Machine learning based short-term travel time prediction: Numerical results and comparative analyses}},
    year = {2021},
    journal = {Sustainability},
    author = {Qiu, Bo and Fan, Wei},
    number = {13},
    month = {7},
    volume = {13},
    publisher = {MDPI AG},
    doi = {10.3390/su13137454},
    issn = {20711050}
}

@inproceedings{Miller2012MiningImpact,
    title = {{Mining traffic incidents to forecast impact}},
    year = {2012},
    booktitle = {UrbComp '12: Proceedings of the ACM SIGKDD International Workshop on Urban Computing},
    author = {Miller, Mahalia and Gupta, Chetan},
    pages = {33--40},
    isbn = {9781450315425},
    doi = {10.1145/2346496.2346502}
}

@article{TavassoliHojati2016ModellingReliability,
    title = {{Modelling the impact of traffic incidents on travel time reliability}},
    year = {2016},
    journal = {Transportation Research Part C: Emerging Technologies},
    author = {Tavassoli Hojati, Ahmad and Ferreira, Luis and Washington, Simon and Charles, Phil and Shobeirinejad, Ameneh},
    month = {4},
    pages = {49--60},
    volume = {65},
    publisher = {Elsevier Ltd},
    doi = {10.1016/j.trc.2015.11.017},
    issn = {0968090X}
}

@article{Chalumuri2014ModellingJapan,
    title = {{Modelling travel time distribution under various uncertainties on Hanshin expressway of Japan}},
    year = {2014},
    journal = {European Transport Research Review},
    author = {Chalumuri, Ravi Sekhar and Yasuo, Asakura},
    number = {},
    month = {3},
    pages = {85--92},
    volume = {6},
    publisher = {Springer Verlag},
    doi = {10.1007/s12544-013-0111-3},
    issn = {18668887}
}

@article{Sullivan1997NewDelays,
    title = {{New Model for Predicting Freeway Incidents and Incident Delays}},
    year = {1997},
    journal = {Journal of Transportation Engineering},
    author = {Sullivan, Edward C},
    number = {4},
    pages = {267--275},
    volume = {123}
}

@article{Kim2005OptimalInformation,
    title = {{Optimal vehicle routing with real-time traffic information}},
    year = {2005},
    journal = {IEEE Transactions on Intelligent Transportation Systems},
    author = {Kim, Seongmoon and Lewis, Mark E. and White III, Chelsea C.},
    number = {2},
    month = {6},
    pages = {178--188},
    volume = {6},
    doi = {10.1109/TITS.2005.848362},
    issn = {15249050}
}

@article{Li2018OverviewPrediction,
    title = {{Overview of traffic incident duration analysis and prediction}},
    year = {2018},
    journal = {European Transport Research Review},
    author = {Li, Ruimin and Pereira, Francisco C. and Ben-Akiva, Moshe E.},
    number = {},
    month = {6},
    pages = {22},
    volume = {10},
    publisher = {Springer Verlag},
    doi = {10.1186/s12544-018-0300-1},
    issn = {18668887}
}

@inproceedings{vanHinsbergen2007ShortModels,
    title = {{Short term traffic prediction models}},
    year = {2007},
    booktitle = {Proceedings of the 14th World Congress on Intelligent Transport Systems (ITS)},
    author = {van Hinsbergen, C P and van Lint, J W and Sanders, F M},
    pages = {},
    publisher = {},
    language = {Undefined/Unknown}
}

@article{Vlahogianni2014Short-termGoing,
    title = {{Short-term traffic forecasting: Where we are and where we're going}},
    year = {2014},
    journal = {Transportation Research Part C: Emerging Technologies},
    author = {Vlahogianni, Eleni I. and Karlaftis, Matthew G. and Golias, John C.},
    pages = {3--19},
    volume = {43},
    publisher = {Elsevier Ltd},
    doi = {10.1016/j.trc.2014.01.005},
    issn = {0968090X}
}

@article{Kim2005StateInformation,
    title = {{State space reduction for non-stationary stochastic shortest path problems with real-time traffic information}},
    year = {2005},
    journal = {IEEE Transactions on Intelligent Transportation Systems},
    author = {Kim, S and Lewis, M E and White III, C C},
    number = {3},
    pages = {273--284},
    volume = {6},
    doi = {https://doi.org/10.1109/TITS.2005.853695}
}

@book{Tijms1986StochasticApproach,
    title = {{Stochastic Modelling and Analysis: A Computational Approach}},
    year = {1986},
    author = {Tijms, Henk C},
    publisher = {John Wiley {\&} Sons, Inc.},
    address = {},
    isbn = {0471909114}
}

@article{Fosgerau2010TheReliability,
    title = {{The value of reliability}},
    year = {2010},
    journal = {Transportation Research Part B: Methodological},
    author = {Fosgerau, Mogens and Karlstr{\"{o}}m, Anders},
    number = {1},
    pages = {38--49},
    volume = {44},
    doi = {https://doi.org/10.1016/j.trb.2009.05.002},
    issn = {0191-2615}
}

@article{Yeon2008TravelChains,
    title = {{Travel time estimation on a freeway using Discrete Time Markov Chains}},
    year = {2008},
    journal = {Transportation Research Part B: Methodological},
    author = {Yeon, Jiyoun and Elefteriadou, Lily and Lawphongpanich, Siriphong},
    number = {4},
    pages = {325--338},
    volume = {42},
    publisher = {Elsevier Ltd},
    doi = {10.1016/j.trb.2007.08.005},
    issn = {01912615}
}

\appendix
\section{Conditional Erlang distribution} \label{appendix:conditionalErlang}

\begin{theorem} \label{thm:conditionalErlang}
For $t \in \mathbb{R}_{> 0}$ and $X \sim \text{Erlang}(k,\mu)$,
the distribution of $X \,|\, X\!>\! t$
is a mixture of Erlang$(j,\mu)$ distributions with
$j=1,\dots,k$.
\end{theorem}
\begin{proof}
For an Erlang$(k,\mu)$ distribution we have:
\[
\tilde{p}(t) := \mathbb{P}(X > t) = \sum_{n = 0}^{k\!-\!1}
\frac{e^{-\mu t}}{n!}(\mu t)^n.
\]
Thus, Newton's Binomium gives that:
\begin{align*}
     \mathbb{P}(X\!>\!t\!+\!s\,|\,X\!>\!t) 
    &= \frac{1}{\tilde{p}(t)}\sum_{n = 0}^{k-1}
    \frac{e^{-\mu(t+s)}}{n!}(\mu(t\!+\!s))^n = 
    \frac{1}{\tilde{p}(t)}\sum_{n = 0}^{k\!-\!1}
    \sum_{j = 0}^n \frac{e^{-\mu(t+s)}}{j!(n\!-\!j)!}(\mu t)^j(\mu s)^{n-j}
    \\&=
    \frac{1}{\tilde{p}(t)}\sum_{j = 0}^{k-1}
    \sum_{n=j}^{k-1} \frac{e^{-\mu(t+s)}}{j!(n\!-\!j)!}(\mu t)^j(\mu s)^{n-j}
    =
    \frac{1}{\tilde{p}(t)}\sum_{j = 0}^{k-1}
    \frac{e^{-\mu t}}{j!}(\mu t)^j 
    \sum_{n=0}^{k\!-\!1\!-\!j} \frac{e^{-\mu s}}{n!}(\mu s)^{n}
    \\ & =
    \frac{1}{\tilde{p}(t)}\sum_{j = 0}^{k-1}
    \frac{e^{-\mu t}}{(k\!-\!1\!-\!j)!}(\mu t)^{k-1-j} 
    \sum_{n=0}^{j} \frac{e^{-\mu s}}{n!}(\mu s)^{n}=\sum_{j = 1}^{k} \tilde{p}_j(t)
    \sum_{n=0}^{j-1} \frac{e^{-\mu s}}{n!}(\mu s)^{n},
\end{align*}
with
\[\tilde{p}_j(t):=\frac{1}{\tilde{p}(t)}
    \frac{e^{-\mu t}}{(k\!-\!j)!}(\mu t)^{k-j}. \]
We conclude that the remaining time of an Erlang$(k,\mu)$ random variable, conditioned on being at least $t$, is
a mixture of Erlang$(j,\mu)$ distributions with
$j=1,\dots,k$; with probability $\tilde p_j(t)$ there are $j$ phases. 
Indeed,
\[
\sum_{j = 1}^k \tilde{p}_j(t)
= \frac{1}{\tilde{p}(t)} \sum_{j = 1}^k 
    \frac{e^{-\mu t}}{(k\!-\!j)!}(\mu t)^{k-j}
= \frac{1}{\tilde{p}(t)} \sum_{j = 0}^{k-1}
    \frac{e^{-\mu t}}{(k\!-\!j\!-\!1)!}(\mu t)^{k-j-1}
= \frac{1}{\tilde{p}(t)}\sum_{j = 0}^{k-1} 
    \frac{e^{-\mu t}}{j!}(\mu t)^{j} = 1.
\]
\end{proof}

\end{document}